\def\year{2022}\relax
\documentclass[letterpaper]{article} % DO NOT CHANGE THIS
\usepackage{aaai22}  % DO NOT CHANGE THIS
\usepackage{times}  % DO NOT CHANGE THIS
\usepackage{helvet}  % DO NOT CHANGE THIS
\usepackage{courier}  % DO NOT CHANGE THIS
\usepackage[hyphens]{url}  % DO NOT CHANGE THIS
\usepackage{graphicx} % DO NOT CHANGE THIS
\usepackage{natbib}  % DO NOT CHANGE THIS AND DO NOT ADD ANY OPTIONS TO IT
\usepackage{caption} % DO NOT CHANGE THIS AND DO NOT ADD ANY OPTIONS TO IT
\DeclareCaptionStyle{ruled}{labelfont=normalfont,labelsep=colon,strut=off} % DO NOT CHANGE THIS
\usepackage{algorithm}
\usepackage{algorithmic}

\usepackage{subfig} % used for sub figures (e.g., communities social networks)
\usepackage{multirow}   %to be used in tables for multiple rows
\usepackage[table]{xcolor}
\usepackage{arydshln} %to be used for horizonatl lines in tables
\usepackage{enumitem}
\usepackage{hyperref}
\usepackage[nointegrals]{wasysym}
\usepackage{amsmath}
\usepackage{amssymb}

\usepackage[show]{notes}
\newcommand{\rp}{\texttt{r/place}}
\newcommand{\ignore}[1]{}
\newcommand{\com}[1]{}
\newcommand{\oren}[1]{\inote[oren]{\textcolor{red}{\bf #1}}}
\newcommand{\avra}[1]{\inote[Avrahami]{\textcolor{blue}{\bf #1}}}

\usepackage{newfloat}
\usepackage{listings}
\floatstyle{ruled}
\newfloat{listing}{tb}{lst}{}
\floatname{listing}{Listing}
\nocopyright
\title{With Flying Colors: \\ Predicting Community Success in Large-scale Collaborative Campaigns}
\author {Abraham Israeli ~~~~~~~~~~~~~~~~ Oren Tsur}
\affiliations {
    Department of Software and Information System Engineering\\
    Ben-Gurion University of the Negev, Israel\\~~~~~~~isabrah@post.bgu.ac.il~~~~~~~~~~~~~~~ orentsur@bgu.ac.il}

\begin{document}
\maketitle
\begin{abstract} 
Online communities develop unique characteristics, establish social norms, and exhibit distinct dynamics among their members. Activity in online communities often results in concrete ``off-line'' actions with a broad societal impact (e.g., political street protests and norms related to sexual misconduct). While community dynamics, information diffusion, and online collaborations have been widely studied in the past two decades, quantitative studies that measure the effectiveness of online communities in promoting their agenda are scarce. In this work, we study the correspondence between the effectiveness of a community, measured by its success level in a competitive online campaign, and the underlying dynamics between its members. To this end, we define a novel task: predicting the success level of online communities in Reddit's \rp~ -- a large-scale distributed experiment that required collaboration between community members.
We consider an array of definitions for success level; each is geared toward different aspects of collaborative achievement. We experiment with several hybrid models, combining various types of features. Our models significantly outperform all baseline models over all definitions of `success level'.
Analysis of the results and the factors that contribute to the success of coordinated campaigns can provide a better understanding of the resilience or the vulnerability of communities to online social threats such as election interference or anti-science trends. We make all data used for this study publicly available for further research.
\end{abstract}

~~~~~~~~{\bf [Accepted for publication at ICWSM 2024]}

\section{Introduction}
\label{sec:intro}
Communities, whether offline or online, play a crucial role in how we establish, perceive, and project our identity \cite{Lewin1947groupd,zachary1977information,ostrom2000collective,mcmillan1986sense,cote1996sociological,olson2009logic}.
The fundamental role of the community, its evolving norms, the dynamics between its members, its organizing principles and collective action are studied for decades, e.g., \cite{lewin1947frontiers,granovetter1973strength,ostrom2000collective,fisher2019science,israeli2022must}, to mention just a few works.

The rise of online social platforms provides a unique opportunity to study phenomena that are associated with online communities organically and at a large scale \cite{melucci1996challenging,lazer2009life}. It was shown that online activity often relates to, or even inspires, coordination off-line, such as support for a social change \cite{hassler2020large}, the financial markets \cite{mancini2022self}, %removed lucchini2021reddit due to lack of space 
street protest \cite{jackson2016ferguson,fisher2019science}, and violent outbursts \cite{capitol2021}. % removed jost2018social due to lack of space

%lately became crucial in the political arena as well. Micro-targeting practices used by firms like Cambridge Analytica for political purposes \cite{ward2018social, hu2020cambridge}, the activity of Russian trolls \cite{jamieson2018cyberwar,benkler2018network,mueller2019report} in the spread of political misinformation and interfering with political processes \cite{bovet2019influence,hanouna2019sharp,grinberg2019fake}, and the well-coordinated storming of the U.S. Capitol on January 2021 \cite{capitol2021} by Parler\footnote{Parler -- a social platform, established on August 2018, mainly attracts alt-right users.} communities are a few examples of such political activity.

The literature on large-scale studies of decentralized community operation and coordination is limited. Furthermore, research providing insights into the factors that contribute to the successful execution of collective actions is scarce.

In this work, we aim to quantify, model, and predict the level of community success in a large-scale online campaign that requires collaboration among community members. Our definition of success deviates from traditional metrics such as the number of registered users or the retention rate of members within the community. Instead, we consider several measures, each capturing a slightly different aspect of the notion of success in a concrete campaign: accounting for the complexity of the campaign objective, the community resources, or the opposition it faces. Furthermore, our prediction models are interpretable, allowing us to analyze the contribution of different factors to the success level. This analysis, in turn, provides novel insights and validates the existing theory.

\begin{figure*}[t]
\normalsize
    \centering
    \subfloat[{$t_0+2$ hours \label{subfig:place2hrs}}]{{\includegraphics[scale=0.1093]{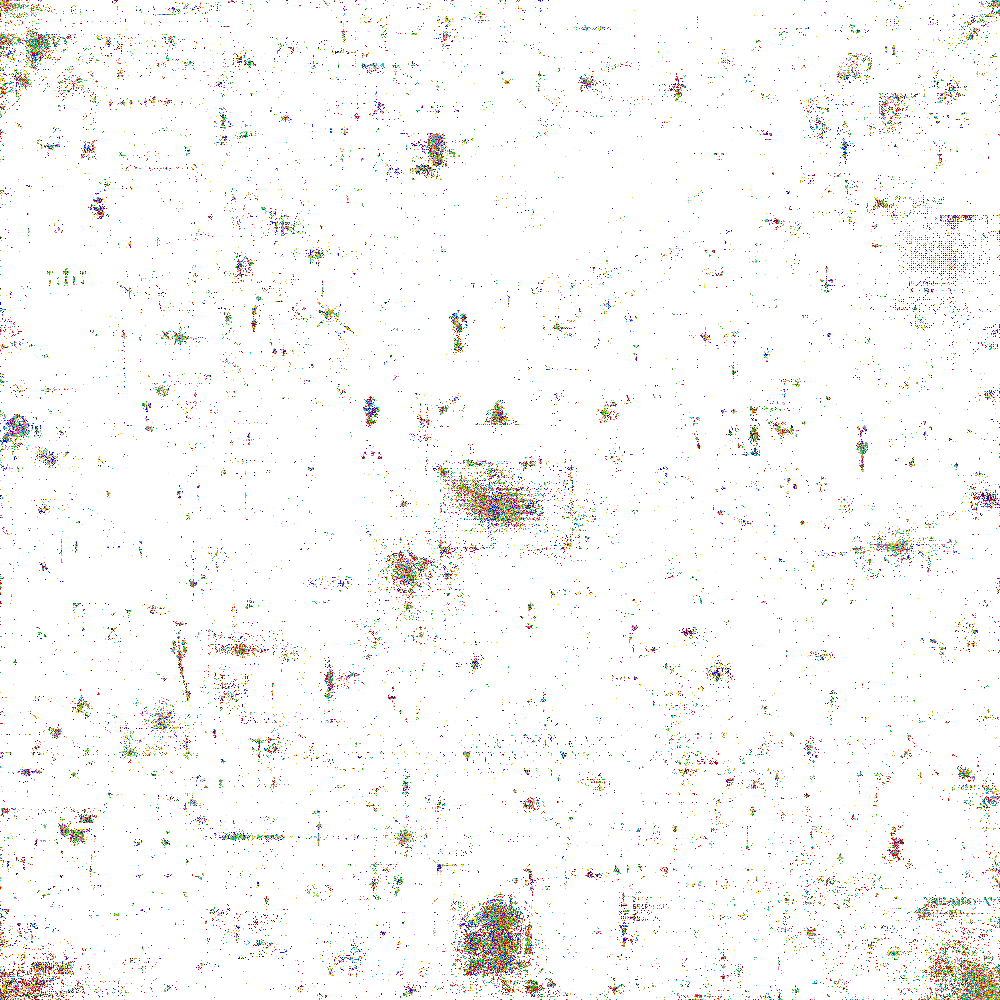}}}
    \qquad
    \subfloat[$t_0+7$ hours 
    \label{subfig:place7hrs}]{{\includegraphics[scale=0.0277]{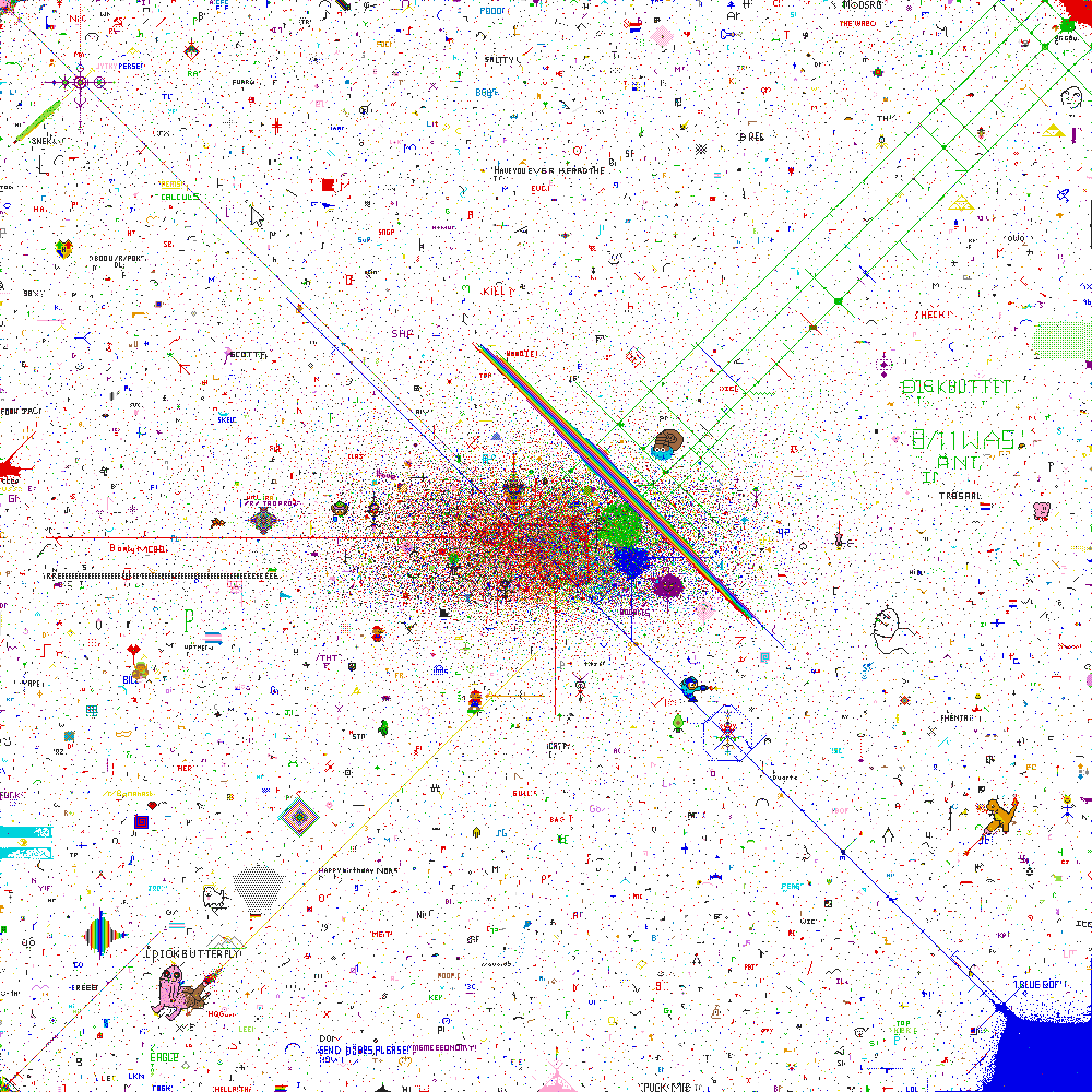}}}
    \qquad
    \subfloat[$t_0+25$ hours\label{subfig:place25hrs} ]{{\includegraphics[scale=0.0277]{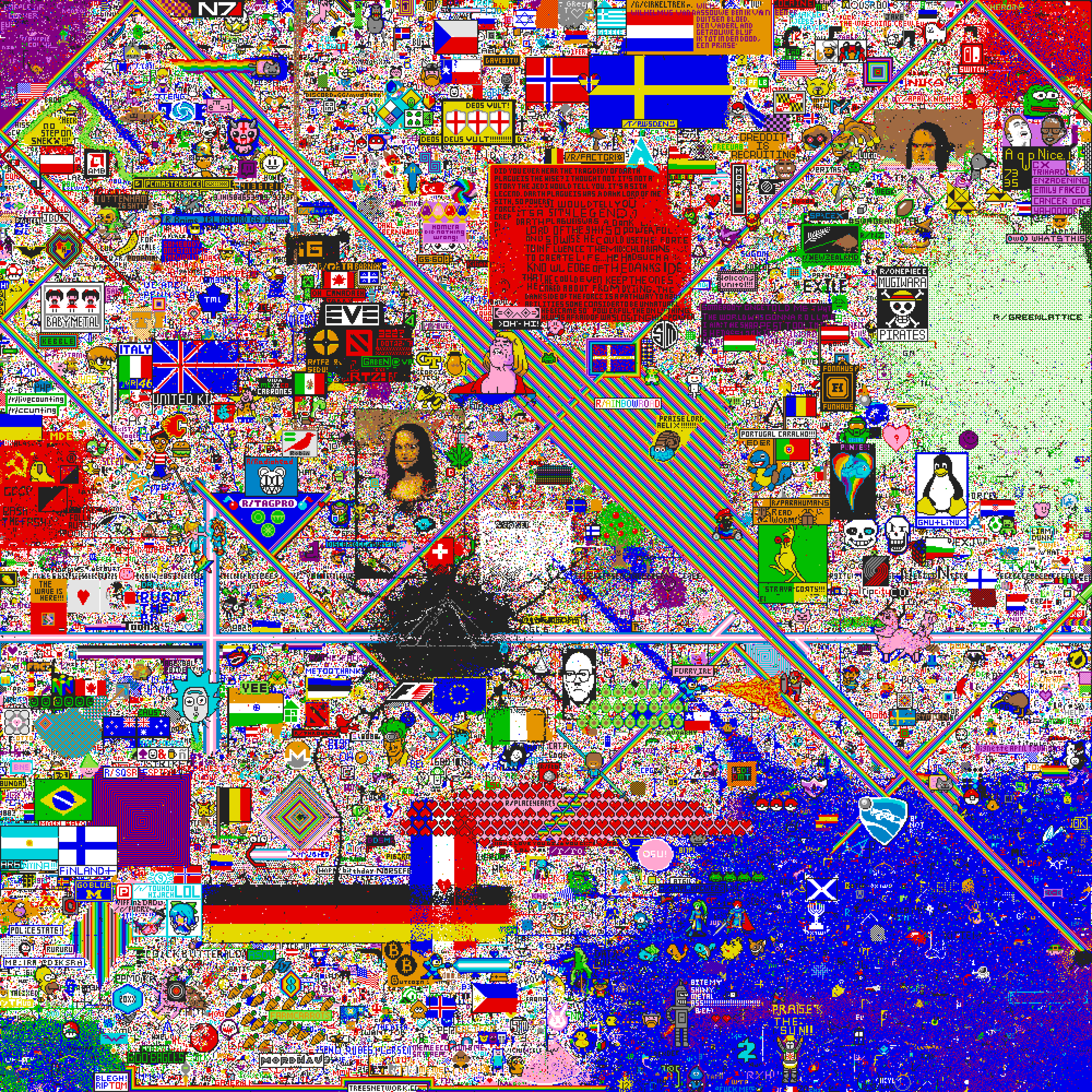}}}
    \qquad
    \subfloat[$t_0+72$ hours (final)\label{subfig:place72hrs} ]{{\includegraphics[scale=0.0277]{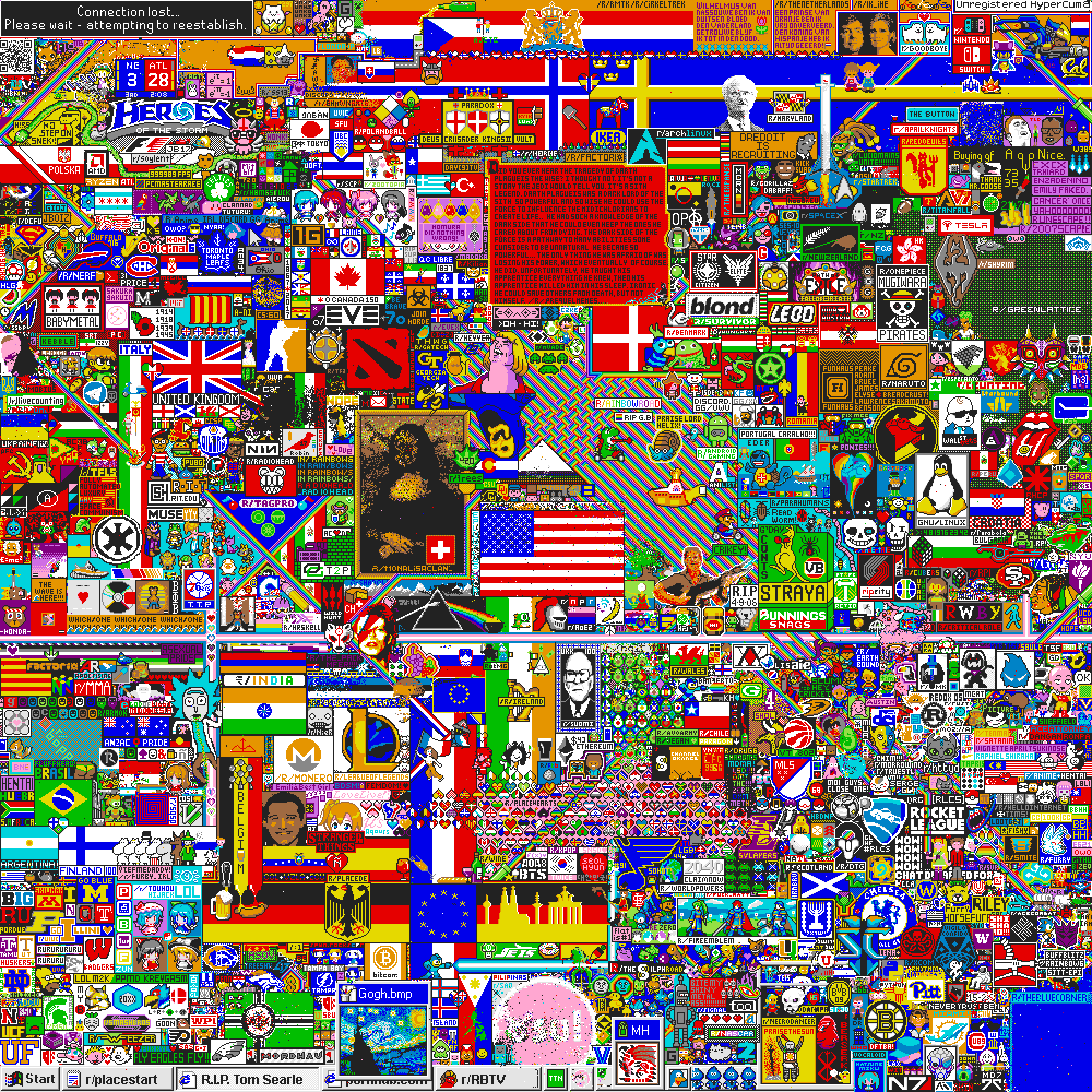}}}
    \qquad
    \hfill \break
    % now the osu VS TBV conflict
    \subfloat[$t_0+24$ hours\label{subfig:osu_TBV24_hours} ]{{\includegraphics[scale=0.0939]{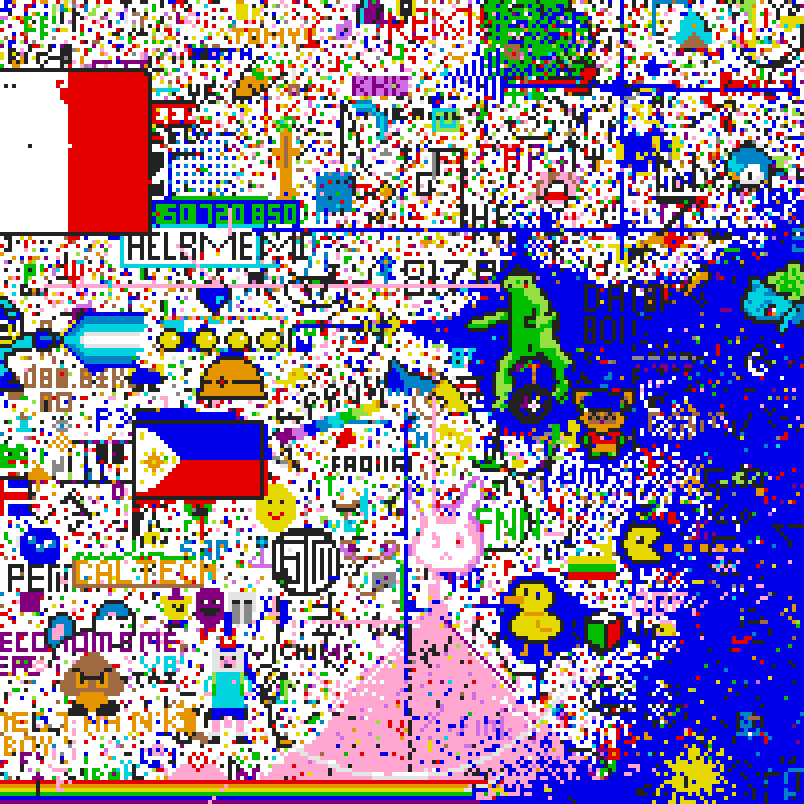}}}
    \quad
    \subfloat[$t_0+28$ hours\label{subfig:osu_TBV28_hours} ]{{\includegraphics[scale=0.0939]{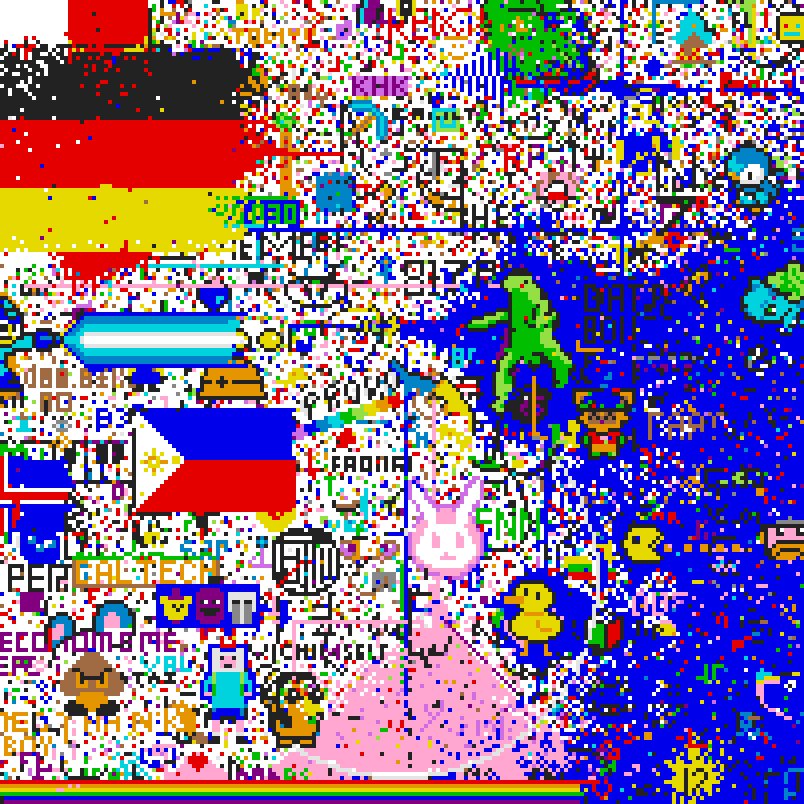}}}
    \quad
    \subfloat[$t_0+39$ hours\label{subfig:osu_TBV39_hours} ]{{\includegraphics[scale=0.0939]{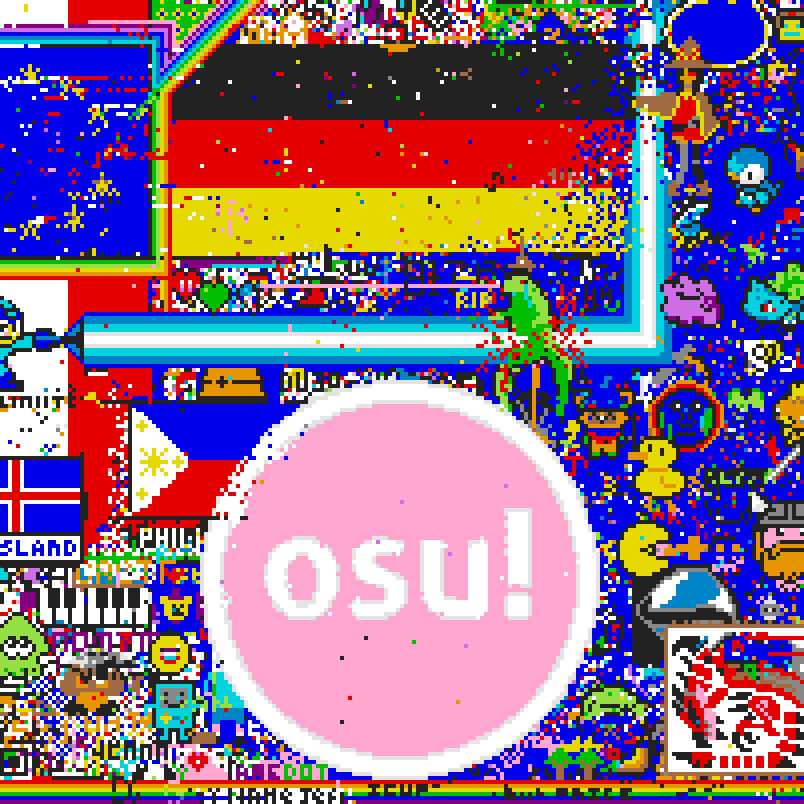}}}
    \quad
    \subfloat[$t_0+42$ hours\label{subfig:osu_TBV42_hours} ]{{\includegraphics[scale=0.0939]{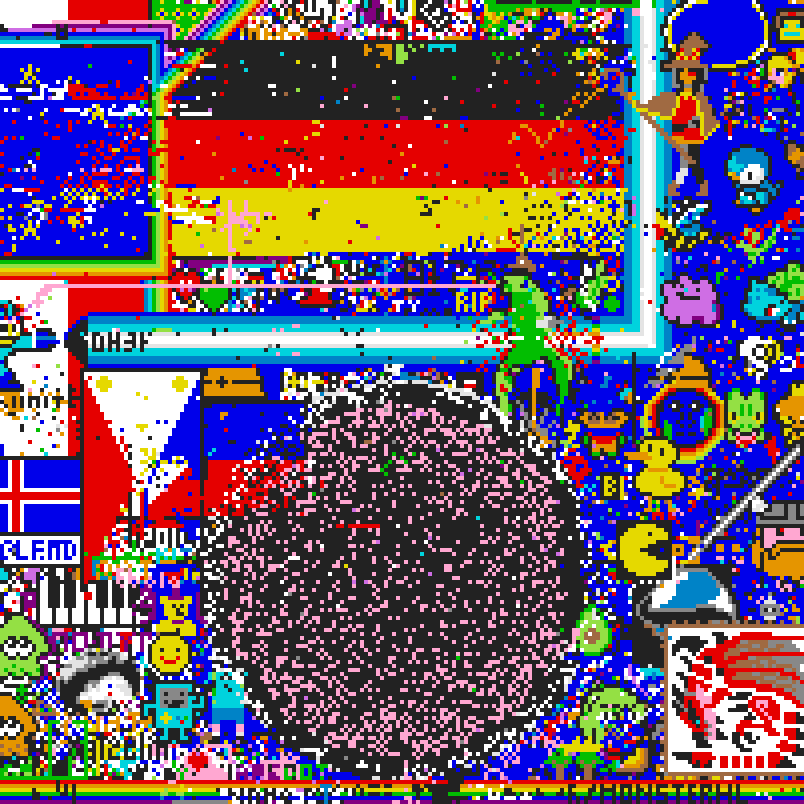}}}
    \quad
    \subfloat[$t_0+68$ hours\label{subfig:osu_TBV68_hours} ]{{\includegraphics[scale=0.0939]{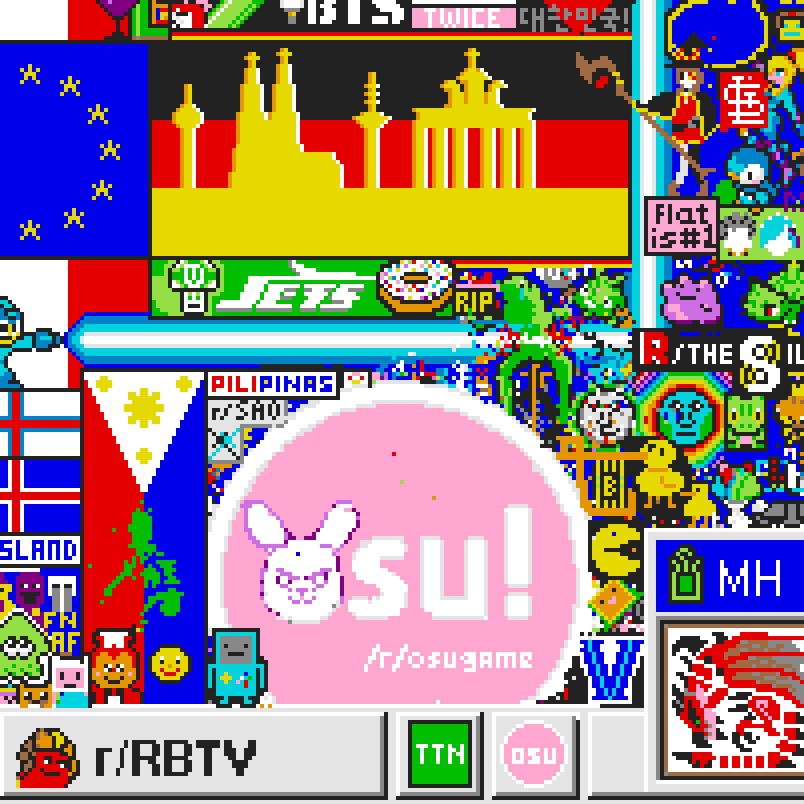}}}
    \quad
    \subfloat[$t_0+72$ hours\label{subfig:osu_TBV72_hours} ]{{\includegraphics[scale=0.0939]{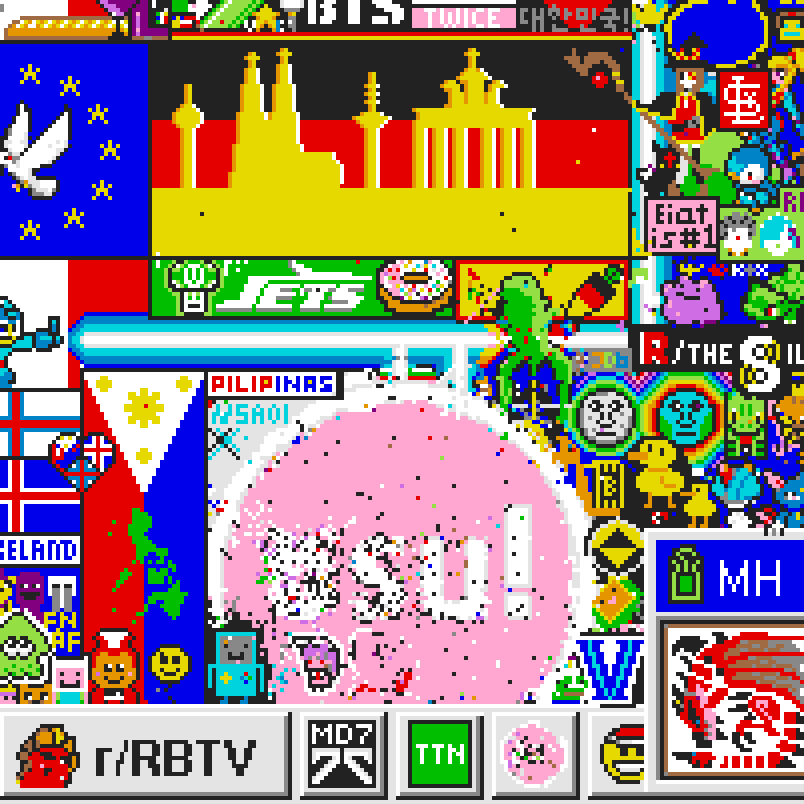}}}
    \hfill \break
    % now the rangers VS pakistan VS Shiv Sena conflict
    \subfloat[{$t_0+54$ hours \label{subfig:rangers_pakistan54hrs}}]{{\includegraphics[scale=0.1445]{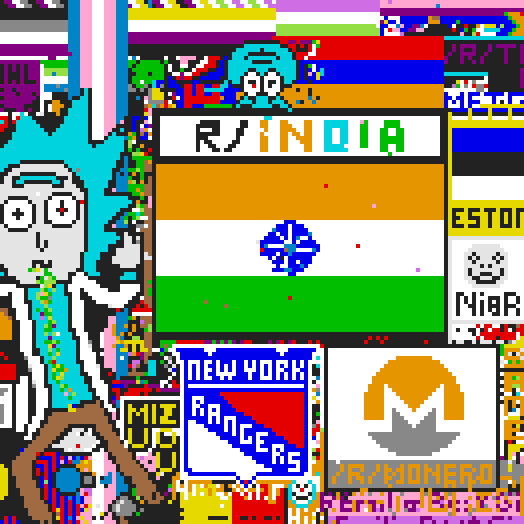}}}
    \quad
    \subfloat[$t_0+55$ hours 
    \label{subfig:rangers_pakistan55hrs}]{{\includegraphics[scale=0.144]{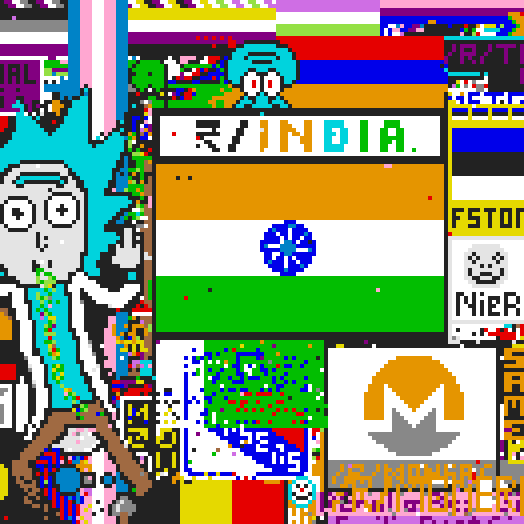}}}
    \quad
    \subfloat[$t_0+60$ hours\label{subfig:rangers_pakistan60hrs} ]{{\includegraphics[scale=0.144]{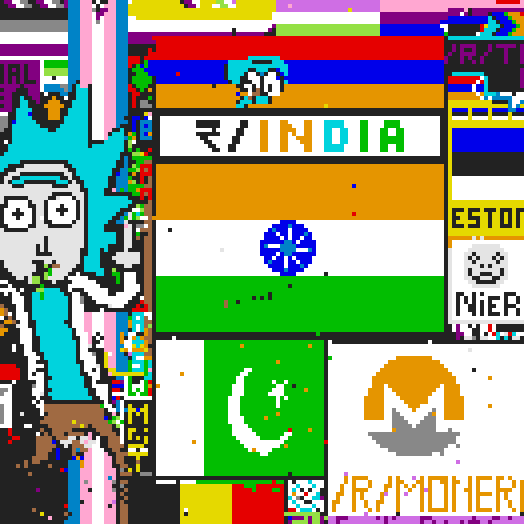}}}
    \quad
    \subfloat[$t_0+66$ hours\label{subfig:rangers_pakistan66hrs} ]{{\includegraphics[scale=0.144]{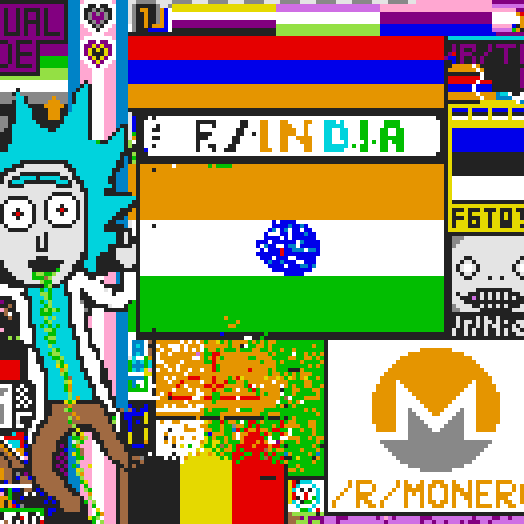}}}
    \quad
    \subfloat[$t_0+69$ hours\label{subfig:rangers_pakistan69hrs} ]{{\includegraphics[scale=0.144]{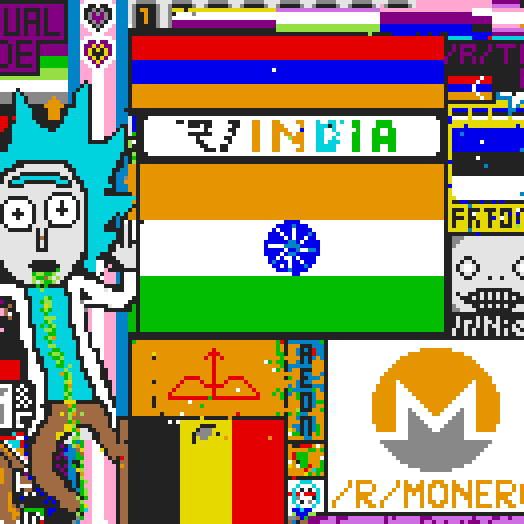}}}
    \quad
    \subfloat[$t_0+72$ hours\label{subfig:rangers_pakistan72hrs} ]{{\includegraphics[scale=0.144]{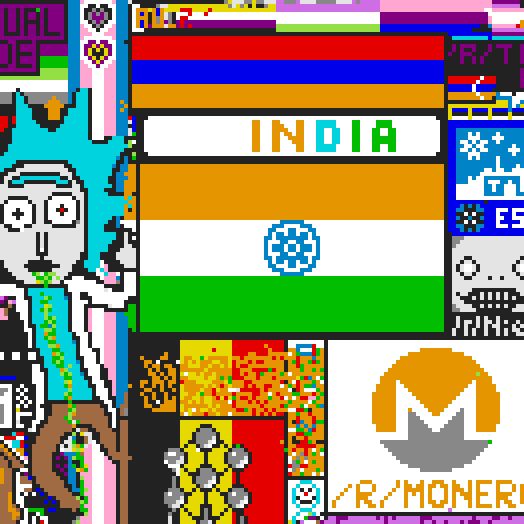}}}
    \quad
    \caption{\small Snapshots of the evolution of the full canvas, and zoom-in on illustrative conflicts. \emph{Top row}: Evolution of the full canvas; \emph{Middle row}: %A conflict region between Germany and France during the first hours of the experiment (upper left in Figures (e)-(f)) until a peace treaty is signed. 
    The logo of the OSU video game is generated by the OSU community, then repeatedly vandalized by TheBlackVoid community; \emph{Bottom row}: The New York Rangers, Pakistan, Shiv Sena, and Belgium clash over the control of the area just under the Indian flag.
    The top row snapshots are taken from \citet{israeli2022must}, and the other snapshots were generated from the archive gallery of the \rp~provided by \citet{pietroalbini2017}. A high-resolution image of the final state of the canvas is available at \href{https://bit.ly/39e1E9a}{https://bit.ly/39e1E9a}.}
    \label{fig:canvas_evolution_conflict_regions}
\end{figure*}

\paragraph{{\bf Reddit and the \rp~experiment}} Reddit\footnote{Reddit website: \href{https://www.reddit.com}{https://www.reddit.com}} is one of the most popular social media worldwide. On average, it attracts more than 430 million active users per month \cite{reddit2022stats} that communicate at over 3.4M forums (as of December 2022), called \emph{subreddits}. In each subreddit, users (\emph{redditors}) can initiate a discussion thread, contribute to a thread, and up/down-vote other posts. Each subreddit constitutes a community that develops informal norms and formal rules.

%Titled as ``the best internet's experiment yet''\footnote{Newsweek article: \href{https://tinyurl.com/49v856a3}{https://tinyurl.com/49v856a3}} by Newsweek, 
The \rp\footnote{\rp~subreddit: \href{https://www.reddit.com/r/place/}{https://www.reddit.com/r/place/}} experiment was launched by Reddit on April Fools’ Day, 2017. A shared white canvas of one million pixels (1000 x 1000) appeared in a new subreddit called r/place. Redditors could select any pixel and change its color. Every change was reflected on the \emph{shared} canvas, thus viewed by all ``participants''.

Once a redditor recolored a pixel, he/she was automatically blocked by the system for some random time (5–20 minutes), effectively preventing any single redditor from having any significant or lasting effect on the canvas. 
%encouraging redditors to collaborate with others -- as mentioned in the r/place subreddit guidelines: ``Individually you can create something. Together you can create something more''. 
The \rp~was not conceived with any specific purpose or goal, thus users were encouraged to do anything in particular. 
To the users' surprise (and dismay?), the canvas was blocked for further manipulation after 72 hours.  

Over the course of the experiment, the canvas was manipulated over 16 million times by 1.2 million Redditors. We refer to \rp~as a naturally occurring large-scale controlled experiment. Figure \ref{fig:canvas_evolution_conflict_regions} (top row) presents four snapshots attesting to the progression of the canvas' state, from its early chaotic state to its final shape -- a diversified collage of complex logos, flags, symbols, and artworks.

\paragraph{Identity, clashes, and success in \rp} In order to provide the appropriate definitions of success in a collaborative campaign, we should first present some of the dynamics that unfolded during the 72 hours of the experiment.

Examining the final state of the canvas (Figure \ref{subfig:place72hrs}), one can observe that most of the artworks are associated with a well-defined identity, e.g., national flags, mascots, emblems of colleges and sport clubs, or gaming communities. Interestingly, a number of  \emph{new} communities were established during the experiment.\footnote{Note that while these organizations are new, only pre-existing users could manipulate the canvas.} The most successful and recognizable ones are the monochromatic efforts -- `The-Blue-Corner' (bottom right in Figures \ref{subfig:place7hrs}-\ref{subfig:place72hrs}), %`The-Green-Lattice' (middle right in Figure \ref{subfig:place25hrs}), 
and `The-Black-Void' (TBV) (middle of Figure \ref{subfig:place25hrs}).

TBV was an organized trolling effort to vandalize the canvas by recruiting many Redditors to expand a black fractal-like shape in multiple regions, overriding other artworks. For example, consider the logo of the OSU video game, created gradually (Figures \ref{subfig:osu_TBV24_hours} -- \ref{subfig:osu_TBV39_hours}), to be ``raided'' by TBV (see Figure \ref{subfig:osu_TBV42_hours}). The OSU logo is later recovered (Figure \ref{subfig:osu_TBV68_hours}) and attacked again, though with limited success (notice the black pixels scattered on the logo in \ref{subfig:osu_TBV72_hours}).

It is important to note that while the declared purpose of TBV's  was pure vandalism, clashes between communities competing for ``real estate'' were common. An example is provided in Figures \ref{subfig:rangers_pakistan54hrs} -- \ref{subfig:rangers_pakistan72hrs} in which The New York Rangers, Pakistan, Shiv Sena (Indian nationalist party), and Belgium clash over the control of the area just under the Indian flag. 
Also note that while the Belgium community managed to successfully expand their flag, dominating the area at the termination of \rp, other communities were not as fortunate, having to relocate or disappear.

The \rp~experiment provides a straightforward and (almost) unmoderated setting in which only coordinated efforts could have any significant impact on the final state. As such, it provides a unique opportunity to study, on a large scale, how decentralized communities coordinate to achieve a common goal in a state of ``emergency'' in which their efforts are hindered and sabotaged by adversaries. Understanding the factors that contribute to the success of coordinated campaigns may shed new light (or validate social theory) regarding the resilience or vulnerability of communities to manipulation campaigns such as election interference or anti-science trends.

\paragraph{Measures of success} While seemingly straightforward, success could be defined and measured in multiple ways. These measures may be correlated to a certain degree, but not identical. The simplest indication of success is binary: whether a community managed or failed to leave any recognizable mark on the canvas. Viewing success through this generic binary lens postulates that a community achieving the placement of a logo of a hundred pixels is as successful as a community achieving a thousand pixels. Another simple measure of success could be the number of pixels a community placed. However, ranking the success level based on pixel counts ignores many other factors that should be accounted for. Some relevant factors are the size of the community, the complexity of the logo, or the demand for the location of the logo on the canvas. 

In this work, we explicitly consider these factors (size, location, complexity) and train prediction models to predict success according to the different measures. A further discussion, concrete examples, and formal definitions of \emph{success} are provided in Section \ref{sec:definitions_of_success}. 

Our prediction models take into account a multifaceted representation of each community, combining linguistic patterns (e.g., vocabulary and distributional semantics), the community social structure (e.g., network embedding), and meta-features (e.g., number of active members and age). The full list of feature types is provided in Section \ref{subsec:characterizing_subreddits}, and the prediction models we consider are described in Section \ref{subsec:prediction_models}. \textcolor{black}{We make all code, annotated data, models, and information for reproducibility of the research publicly available on the project's GitHub repository.}\footnote{\href{https://github.com}{Explicit link will be available soon.}}

Finally, we analyze the results and discuss the contribution of various input representations and community features to the success of a community, accounting for the different definitions of success. Moreover, we consider two modes of community representation: (i) Using data that was generated only \emph{prior} to the experiment, and (ii) Representation based on data that was generated only \emph{during} the 72 hours of \rp. These two modes allow us to further explore whether communities that quickly adapt to the ``state of emergency'' perform better. Results and analysis are presented in Section \ref{sec:res_and_analysis}. 

\ignore{
    Using the new success measures, we develop predictive multimodal algorithms based on hundreds of Reddit communities participating in \rp. We integrate multiple signals, ranging from the language used to the community structure. We use the complex signals \emph{preceding} \rp~as well as the ones recorded \emph{in the course} of the experiment, to predict the success level of communities in the game.
    
    Lastly, we analyze the predictive models using explanatory tools \cite{lundberg2017unified}. By that, we aim to gain a deeper understanding of the factors that contribute to a community's success in an online collaborative campaign, as well as identify differences between the various success measures. \textcolor{black}{We make all code, annotated data, models, and required information for reproducibility of the research publicly available on the project's GitHub repository.}\footnote{\href{https://github.com}{Explicit link will be available soon.}}
     
    The remainder of the paper is organized as follows: in Section \ref{sec:data}, we describe the data we use in detail and specify the different prediction tasks. In Section \ref{sec:exp_setup} we present our computational approach, the various algorithms we use, and the experimental setting. In Section \ref{sec:res_and_analysis} we present the results, an error analysis, and give additional insights into the obtained results. In Section \ref{sec:related_work}, we overview related papers to our research. Lastly, in Section \ref{sec:closure} we summarize our work and suggest future research directions.
} %end ignore

% to create this figure, you should run the notebook code located under: /sise/home/isabrah/reddit_canvas/reddit_project_with_yalla_cluster/reddit-tools/r_place_success_analysis/notebooks
% (and it will yield the png file called "success_against_sr_size.png") 
 \begin{figure*}[th]
  \centering
    \begin{tabular}{c@{}}
        \includegraphics[scale=0.5]{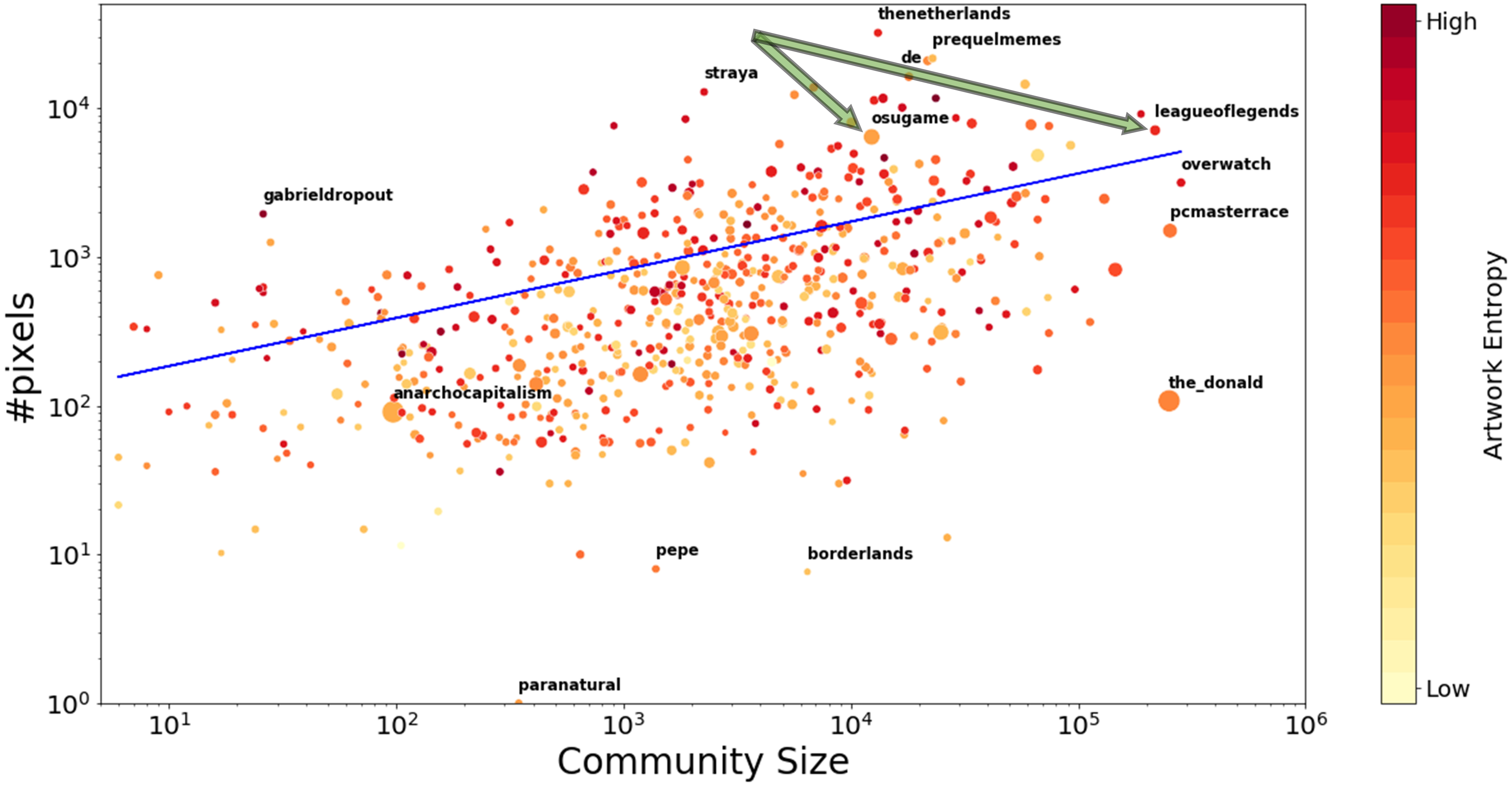}
    \end{tabular}
  \caption{Success level vs. community size. Each point represents a participant community in the \rp~experiment. Highlight communities are explicitly named. The size of each point represents how popular the region in which the community acted was. Colors emphasize the complexity of the drawn artwork, measured by the entropy. The blue line is a linear fit to the data. We use the log scale on both axes for better visibility of the figure.}
  \label{fig:success_against_sr_size}
\end{figure*}

\section{Definitions of Success}
\label{sec:definitions_of_success}
As mentioned in the Introduction, the definition of community success (in a campaign) is not straightforward. \citet{cunha2019all} identified a number of ways to measure the success of a community (e.g., retention rate, growth of membership). However, these measures of success relate to the general state of online communities rather than to the successful coordination toward a specific goal or in mitigating a specific threat. %Through a series of experiments, they show that communities' success has to be measured using multiple measurements, each capturing a different facet of success. 

In this section, we discuss some contextual factors and propose five definitions by which success could be measured. These definitions would serve to assign labels (one binary and four continuous) to be predicted. 

\paragraph{Leaving a mark (binary)} The simplest indication of success is binary:
whether a community managed or failed to leave any recognizable mark on the canvas \textcolor{black}{by the end of the experiment}. This concept of success is demonstrated in the bottom row in Figure \ref{fig:canvas_evolution_conflict_regions}, where the New York Rangers, Pakistan, and Shiv Sena failed to leave a mark as they were eventually overridden by the successful Belgium community.

This naive approach to success is problematic as it postulates that a small yet recognizable logo, of a few pixels, indicates the same success level of a community achieving a much larger logo. We, therefore, wish to consider a success measure that allows ranking, reflecting a success level.

\paragraph{Continuous measures of success} Looking at the \emph{number of pixels} a community has managed to place is a well-defined measure that allows ranking of the success levels. However, this crude measure forgoes many factors that should be considered, marking even smaller logos as a great success.  Some of these factors are: \emph{community size} -- small communities are expected, a-priori, to place fewer pixels; The \emph{complexity} of the campaign objective -- logos with high entropy are harder to coordinate and maintain; \emph{Shape} -- longer ``borders'' are harder to protect (and clashes may erupt in multiple fronts); \emph{Location} -- some areas of the canvas are at a higher demand, thus harder to maintain and protect. 

The impact of these factors on the definition of success is illustrated in Figure \ref{fig:success_against_sr_size}. The regression line (blue) indicates a positive correlation between the size of the community and the size of the logo (number of pixels placed). However, consider the following two gaming communities: r/osuGame and r/LeagueOfLegends (marked by the green arrows in the figure). Both communities allocated about the same number of pixels (6421 and 7114, respectively). However, the number of community members in r/LeagueOfLegends is more than an order of magnitude larger than that of r/osuGame (1.96M and 75.2K, respectively), suggesting that OSU is more successful. On the other hand, the entropy of the LeagueOfLegends logo is much higher than the entropy of OSU (indicated by the darker color), requiring more effort to create and maintain compared to the simplicity of OSU. 

Finally, notice that the area of the OSU logo was in much higher demand compared to that of LeagueOfLegends (indicated by the size of the marker in Figure \ref{fig:success_against_sr_size}), suggesting that although similar in size, leaving a mark of this magnitude in the face of a fiercer opposition is a more impressive accomplishment. 

%end ignore

We, therefore, define four measures of success,  each provides a continuous success score that would be used as non-categorical labels -- all are based on the number of pixels in a logo at the final state, factored by some function $\gamma_k$ to account for different contexts. The success score of a community $c^i$ with respect to $k$ is therefore given by:
\[s_k(i) = \gamma_k(l^i_{\#pixels})\]
where, $l^i_{\#pixels}$ denotes the number of pixels in the logo $l^i$ created by $c^i$, and $k \in \{\phi,|c|,p,d,H\}$ indicates the term by which $l^i_{\#pixels}$ should be factored: 
\begin{itemize}
    \item No factorization ($\phi$).
    \item {\bf Community size} ($|c^i|$): the number users registered as members of community $c^i$.
    \item {\bf Location popularity} ($p^i$): the \emph{total} number of pixel allocations on the area defined by $l^i$, divided by $l^i_{\#pixels}$. 
    \item {\bf Diameter} ($d^i$): the maximal Manhattan distance between two pixels in $l^i$.
    \item {\bf Entropy} ($H^i$): entropy of the distribution of colors in $l^i$.
\end{itemize}

%All $s_k(i)$ values lay in the $[0.1,1]$ range where $1$ is assigned to the most successful community (according the respective $k$). 

%For example, considering two logos $l^i$ and $l^j$ produced by communities $c^i$ and $c^j$, and given that: (i) $l^i_{\#pixels}=l^j_{\#pixels} = \delta$;  (ii) $|c^i|<|c^j|$; (iii) $p^i<p^j$; (iv) $d^i < d^j$; and (v) $H^i > H^j$, we compute the success score and obtain the following relative success ranking: \\
%$s_\phi(i) = s_\phi(j)$, while $s_{\{|c|,H\}}(i) > s_{\{|c|,H\}}(j)$, and  $s_{\{p,d\}}(i) < s_{\{p,d\}}(j)$

\begin{figure}
\normalsize
    \centering
    \subfloat[\label{subfig:black_plus_artwork}]{{\includegraphics[scale=0.18]{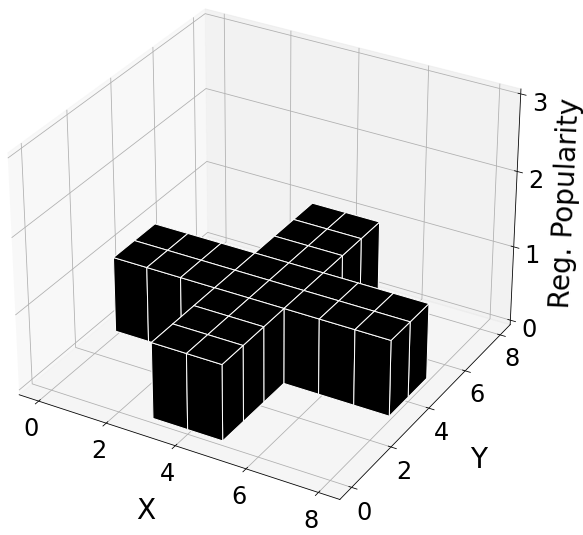}}}
    \qquad
    \subfloat[ 
    \label{subfig:black_red_plus_artwork}]{{\includegraphics[scale=0.18]{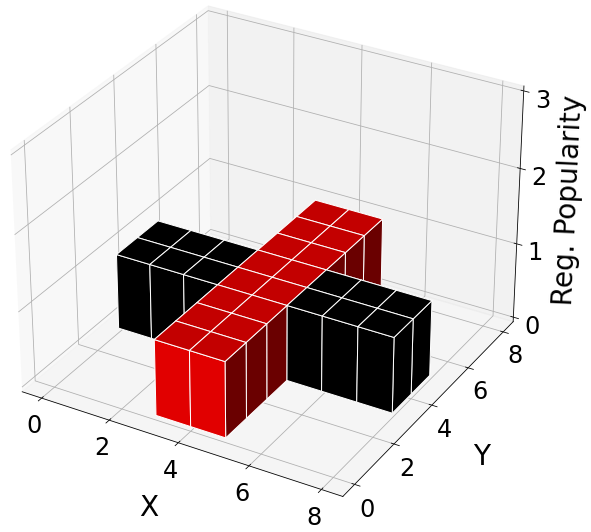}}}
    \qquad
    \subfloat[\label{subfig:black_red_long_artwork}]{{\includegraphics[scale=0.18]{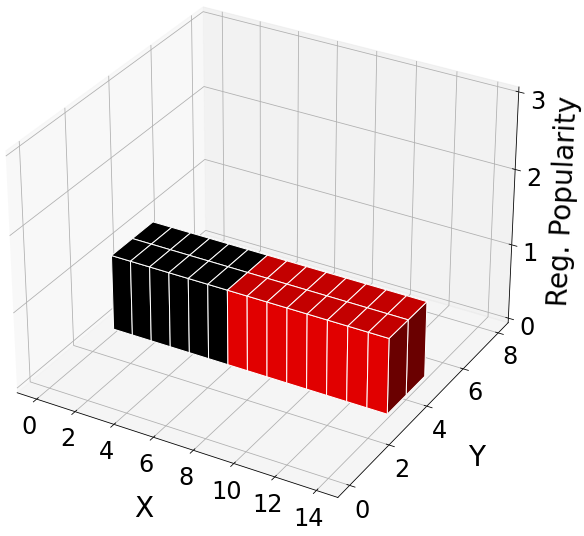}}}
    \qquad
    \subfloat[\label{subfig:black_red_plus_artwork_high_popularity} ]{{\includegraphics[scale=0.18]{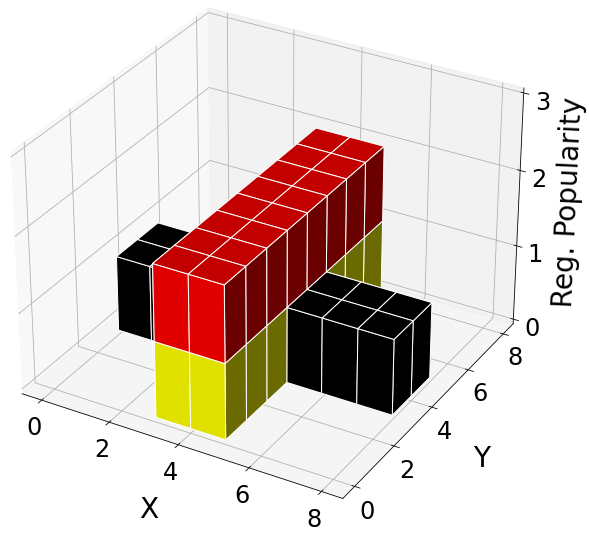}}}
\caption{Artwork examples. On the x-y axis, two-dimensional artworks are presented. The z-axis emphasizes how popular each pixel in the artwork was. All four artworks have 28 allocated pixels.}
  \label{fig:dummy_artworks}
\end{figure}

To guide the eye we illustrate these success measures through four simple toy examples in Figure \ref{fig:dummy_artworks}. All four Figures (\ref{subfig:black_plus_artwork}- \ref{subfig:black_red_plus_artwork_high_popularity}) are made of 28 pixels (unit cubes), projected on the X-Y plane. However, while having an identical shape, the entropy of  Figure \ref{subfig:black_red_plus_artwork} is higher than that of Figure \ref{subfig:black_plus_artwork}, the diameter of Figure \ref{subfig:black_red_long_artwork} is longer than the other logos (14 vs. 8). While the entropy of \ref{subfig:black_red_plus_artwork_high_popularity} is equal to that of \ref{subfig:black_plus_artwork}, the area popularity of Figure \ref{subfig:black_red_plus_artwork_high_popularity} is greater than the popularity of the other logos, as some of the 28 pixels were manipulated twice (presumably the yellow tiles were placed by members of one community, then overridden by another community to create the red-black plus-shaped logo).

Formally, taking the naive approach -- all Figures are equally successful: $s_\phi(a) = s_\phi(b) = s_\phi(c) = s_\phi(d)=28$. 
Taking the diameter into account we obtain the following order $s_d(c) > s_d(a) = s_d(b) = s_d(d)$, while taking the complexity of the logo into account we get $s_H(b) > s_H(d) > s_H(c) > s_H(a)$, and considering the popularity of (demand for) the location of the logo we get $s_p(d) > s_p(a) = s_p(b) = s_p(c)$. 

It is important to note that we find that the diameter positively correlates with the circumference, and therefore it serves as a simple approximation for the number of potential border clashes. Similarly, the definitions of popularity and complexity are only measurable ways that approximate potential dynamics on the canvas.

%end ignore

% a figure showing dummy artworks to explain how our measures are calculated

\section{Data}
\label{sec:data}
\paragraph{Communities}This work takes interest only in the communities that took an active part in the \rp~experiment. While a list of 1231 communities was compiled and shared by \citet{israeli2022must}, we find that due to their use of heuristic filters, the list includes some communities that only discussed \rp~without participating. We manually verified and filtered all communities in their list, obtaining a subset of 997 well-defined communities that took part in the \rp~experiment. 

For each community we obtained a number of meta features (e.g., age and size $|c^i|$) through Reddit's API\footnote{\href{https://www.reddit.com/dev/api/}{https://www.reddit.com/dev/api/}} as well as all the posts, comments and up/down votes during the 72 hours of the experiment and the three months preceding it. These data are used to generate multifaceted representations of each of the communities, which in turn are used as the input for the prediction models (see Section \ref{sec:exp_setup}).

We assigned each community with five gold labels -- one binary and four continuous, each label corresponds to a success measure, as defined in Section \ref{sec:definitions_of_success}.

\paragraph{Binary Gold Labels} The binary labels were obtained by using the \emph{Place-Atlas}\footnote{Place-Atlas: \href{https://draemm.li/various/place-atlas}{https://draemm.li/various/place-atlas}} resource  for manual annotation. We find that 331 communities (33\%) of the participating communities failed to leave a recognizable mark on the  canvas. 

\paragraph{Continuous Gold Labels} The four continuous labels for a community $c^i$ depend on the size of the logo ($l^i_{\#pixels}$), the community size ($|c^i|$), the popularity of the location ($p^i$), the diameter of the logo ($d^i$), and the entropy of the logo ($H^i$). 
Each logo ($l^i$) was identified using the Place-Atlas, from which we directly derived $l^i_{\#pixels}$, and computed $d^i$ and $H^i$. The $|c^i|$ and $p^i$ values were extracted from data shared by Reddit\footnote{\rp~published data: \href{https://tinyurl.com/4ewtwu8w}{https://tinyurl.com/4ewtwu8w}.} and matched with the Place-Atlas data.

%Some communities have multiple artworks scattered on the canvas (e.g., r/france). We made sure to aggregate the values\footnote{The entropy was computed over all artworks by $c^i$.}. We also ensured that overlapping artworks on the Place-Atlas were not counted more than once.

%\oren{center equations. it is ugly this way.}
Finally, the label values for community $c^i$ are given by:\\
\begin{align*}
    &s_{|c|}(i) \triangleq l^i_{\#pixels} \cdot max\{1-\rho( |c^i|), \alpha\}\\
    &s_p(i) \triangleq l^i_{\#pixels} \cdot max\{\rho( p^i), \alpha\}\\
    &s_d(i) \triangleq l^i_{\#pixels} \cdot max\{\rho( d^i), \alpha\}\\
    &s_H(i) \triangleq l^i_{\#pixels} \cdot max\{\rho( H^i), \alpha\}\\
\end{align*}
where $\rho$ denotes the percentile of the value, divided by 100,\footnote{That is if $|c^i|$ is in the 13th percentile, $1-\rho(|c^i|)=0.87$.} and $\alpha=0.1$.\footnote{This hyper-parameter is used to prevent the diminishing of the success score. We find 0.1 to be adequate, although other small values could be used to the same effect.}

\begin{table}
\centering
\footnotesize
{
    \begin{tabular}{l@{ }|c@{\quad}c@{\quad}c@{\quad}c@{}}
      &$s_{|c|}$ & $s_p$ & $s_d$ & $s_H$\\[2pt]\hline
      $s_\phi$ & 0.15 ; 0.41 & 0.06 ; 0.25 & 0.88 ; 0.89 & 0.2 ; 0.24\\[2pt]
      $s_{|c|}$  &  1.0 & 0.16 ; 0.18 & 0.18 ; 0.37 & 0.07 ; 0.09\\[2pt]
      $s_p$ &  & 1.0 & 0.07 ; 0.22 & -0.07 ; -0.01\\[2pt]
      $s_d$ &  &  & 1.0 & 0.22 ; 0.17\\[2pt]
    \end{tabular}
\caption{\small Correlations between the different measures of success. $s_\phi$ is the success rank based on the number of pixels with no accounting for any relevant factor. Value pairs indicate the Pearson (first value) and the Spearman (second value) correlations.}
\label{table:success_label_correlation}
}
\end{table}
 
Table \ref{table:success_label_correlation} presents the correlation between the four measures (and the naive $s_\phi(i) = l^i_{\#pixels}$). We observe a positive, though low, correlation between the measures. The correlation values validate our intuition that the sheer number of pixels should be discounted and that each type of label represents a different nuance of the notion of success.

\paragraph{Temporal Datasets} One of the primary goals of this work is to predict the level of success of a community based on a multifaceted representation that reflects different characteristics of the community. We hypothesize that specific traits and characteristics 
 can explain the success or failure of a community to rally for a cause or against an emerging threat. Furthermore, we wish to test whether these characteristics are exhibited in the community's everyday, mundane, activity, or emerge at a time of crisis. In order to test this, we split the data into two distinct datasets: (i) the data generated during three months \emph{before} \rp~(BP: Before Place), and (ii) the data generated \emph{during} the 72 hours of \rp~(DP: During Place). General statistics describing  $BP$ and $DP$ are presented in Table \ref{table:data_stats}. 
 %Throughout our experiments, we evaluated all prediction models independently for $BP$ and for $DP$.

\begin{table*}
\centering
{\normalsize %defining the text as normal
    \begin{tabular}{l@{ }|c@{\quad}c@{\quad}c@{\quad}c@{\quad}|c@{\quad}c@{\quad}c@{\quad}c@{\quad}}
      \multicolumn{1}{c}{} & \multicolumn{4}{c}{\boldmath{$BP$ (Before Place, 3 months)}} & \multicolumn{4}{c}{\boldmath{$DP$ (During Place, 3 days)}}\\[2pt]
      {} & {Total} & {Mean} & {Median} & {STD} & {Total} & {Mean} & {Median} & {STD}\\[2pt]
      \hline\rule{0pt}{12pt}Active Users	&8.61M  &17.26K   &2.25K  &272.95K  &1.1M   &2.21K    &300    &34.91K\\[2pt]
        Submissions	    &8.4M  &16.84K   &1.79K  &268.68K  &352.4K   &706.1    &89    &11.23K\\[2pt]
        Comments	    &121.65M  &243.79K   &18.31K  &3.87M  &4.7M   &9.43K    &823    &149.43K\\[2pt]
        Tokens       	&3840.8M  &7.7M   &619.17K  &121.93M  &14.02M   &28.09K    &3.08K    &444.86K\\[2pt]
        %Upvotes     	&1383.94M  &2.77M   &126.48K  &45.15M  &57.67M   &115.58K    &5.23K    &1.86M\\[2pt]
    %\hline
    \end{tabular}
}
\caption{Data statistics. $BP$ spans over the period before \rp, while $DP$ spans over the time during \rp. The mean, median, and standard deviation are calculated over the subreddits.}
\label{table:data_stats}
\end{table*}

\section{Experimental Setting}
\label{sec:exp_setup}

\subsection{Community Representation}
\label{subsec:characterizing_subreddits}
Communities are multifaceted and can be characterized from different perspectives. To this end, we represent Reddit communities by features of four general types: (i) Textual features, (ii) Meta-features, (iii) Network features, and (iv) Network Embeddings. We calculated each feature described below over $BP$ and over $DP$ independently.

\paragraph{Textual representations} We normalize the textual data by lower casing, tokenization, removing punctuation, and converting full URL addresses to their domain name only. %(e.g., www.youtube.com/XYZ $\rightarrow$ www.youtube.com). 
We experiment with three types of textual features: (i) Bag-of-words features: We use the {\em TF-IDF} score \cite{salton1986introduction} per token. Using bigrams/trigrams did not yield any improvement, so we report all BOW results for the unigram setting only. (ii) LIWC categories: The Linguistic Inquiry and Word Counts (LIWC) dictionary is used to assign words to cognitive and emotional categories \cite{pennebaker2001linguistic}. A vector of LIWC categories represents utterances -– each entry reflects the weight of the corresponding LIWC category in that text. We aggregate all utterances found in the community discussions to represent each community in a single LIWC feature vector, capturing the ``vibe'' of a community. (iii) Raw text: the community's submissions are concatenated and separated with the [SEP] token to fine-tune a BERT model.  %\footnote{Due to the high number of submissions in specific communities, we restrict the number of submissions per community to 10K.}

\paragraph{Meta-features} Each subreddit can be represented by a series of meta-features. For example, the number of users subscribed to it, the average number of posts per day, the average number of up/down votes per post, the age of the community (days since its creation), etc. We use a total of 25 meta-features per subreddit.

\paragraph{Network features} A community can be characterized by the patterns of communication between its members. These interaction patterns could be thought of as a social network in which a direct reply by user $u$ to a post by user $v$ constitutes a directed edge $u \rightarrow v$. These networks provide another perspective on the organizational principles of a community and the dynamics between its members.
In total, we use 32 network statistics as features (e.g., \#nodes, \#edges, avg. and std. of various centrality measures, \#triangles).%\footnote{The list of all meta and network features, together with a short description of each, is provided in Table \ref{table:meta_and_network_feautures_description} in Appendix \ref{sec:appendix_meta_and_network_features}.}

\paragraph{Community Embeddings} While the network features described above were used by \citet{israeli2022must}, we find this approach naive. We, therefore, consider three stronger representations of the social graph:
\begin{itemize}
\item SNAP embeddings: Pretrained embeddings of 51.2K Reddit communities, including \emph{all} the communities in our data are shared as part of SNAP \cite{kumar2019predicting}. \textcolor{black}{The embeddings are generated based on data from Jan 2014 to April 2017. The dimension of the node embeddings is 300.} The SNAP embeddings are not computed on the $BP$ and $DP$ datasets. 
%\footnote{SNAP dataset: \href{https://tinyurl.com/58pftm9k}{https://tinyurl.com/58pftm9k}}
%Two subreddit embeddings are similar if the users who post in them are similar.
\item Community2vec: We use the community embeddings algorithm \cite{martin2017community2vec} on $BP$ and $DP$ (independently), obtaining embedding vectors with a dimension of 100. %The embeddings are based on the intersection between members of different communities. 
\item Graph2vec: We use the graph social structure of each subreddit to train \emph{an unsupervised} graph-level embeddings using the InfoGraph algorithm \cite{sun2019infograph}. We use the implementation suggested by \citet{liu2021dig} with a learning rate of 0.001, trained over five epochs to yield an embedding vector of size 100 per community.
\end{itemize}

\subsection{Prediction Models}
\label{subsec:prediction_models}
%\oren{need to add citations for the models. also - did you use Delvin's BERT or another implementation like roberta?}
In predicting the binary success labels cast the problem as a binary classification task. For the continuous labels, we formulate the problem as a regression task. 
We experiment with an array of algorithms ranging from simple logistic/linear regression \cite{searle2016linear} to gradient-boosted trees \cite{friedman2002stochastic}, feed-forward neural networks, and transformers \cite{vaswani2017attention}. We use a deviance loss function for the GBT and the Random-Forest algorithms, while a binary log-loss is used in fine-tuning the BERT model. 
%Due to space constraint, we report only on the best-performing algorithm among the various type of models tested. 

\subsection{Experimental Settings}
\label{subsec:exp_setting}
Each subreddit is represented by an array of features as described above (Section \ref{subsec:characterizing_subreddits}), derived separately from the $BP$ and $DP$ data. 
%Note that we experiment with $BP$ and $DP$ independently as some of the communities were generated for \rp~only, hence they do not exist in $BP$.
We execute all algorithms in an ablation manner, in order to evaluate the contribution of the different feature types. Precision, Recall, F1-score, and AUC scores are reported for the binary setting. We report the RMSE and the adjusted R-square in the regression setting. The classification algorithms optimize the F1-score as precision and recall are equally important, given the task definition. The regression algorithms optimize the squared error.
We evaluate all algorithms and settings using stratified 5-fold cross-validation. Neural architectures are restricted to a maximum of ten epochs with early stopping.

%We use the sklearn \cite{pedregosa2011scikit} implementation of the GBT and the Random-Forest. We use PyTorch packages \cite{paszke2019pytorch} for building the neural architectures, due to the dynamic construction of the computation graph which is highly efficient given the high variance in number and length of textual data. We make the code developed as part of the research and all data that were collected public in our project's GitHub repository.\footnote{Link is removed to preserve anonymity.}

\section{Results and Analysis}
\label{sec:res_and_analysis}
We compare our results with two intuitive univariate linear models: one uses the size of the community as the independent variable, and the other uses the community's age (days since creation, counting back from 31/3/2017).

%Our findings indicate that the utilization of BERT models \cite{devlin2018bert} is inferior compared to alternative models. We further discuss these results in Section \ref{sec:discussion}.

\begin{table}
\centering
\footnotesize    %defining the text as normal size
    {
    \begin{tabular}{l@{\quad}c@{\quad}|c@{\quad}|c@{\quad}@{\quad}c@{}}
      {Succ.} & {Dataset} & {Features} & {RMSE ($\downarrow$)} & {Adj. $R^2$ ($\uparrow$)}\\[2pt]
      \hline\rule{0pt}{12pt}
      % first gold label - pixels + community size
        \multirow{4}{*}{\rotatebox[origin=c]{360}{\footnotesize $s_{|c|}$}} &
        \multirow{2}{*}{\vtop{\hbox{\strut Ext.-}\hbox{\strut Baseline}}}
        & $|c|$         &0.558$\pm$0.03     &0.045$\pm$0.045     \\[2pt]
        & & Age   &0.576$\pm$0.039     &-0.015$\pm$0.007     \\[2pt]

    \cdashline{2-5} \rule{0pt}{12pt}
    %\cline{2-5}\cline{2-5}\rule{0pt}{12pt}
    & \multirow{1}{*}{$BP$}
        & Network    &0.535$\pm$0.042     &0.124$\pm$0.063     \\[2pt]

    % end of $BP$ models - moving to $DP$
    %\cdashline{2-5}
    & \multirow{1}{*}{$DP$}
        & All   &0.519$\pm$0.048    &0.291$\pm$0.021     \\[2pt]
    % next gold label (pixels + popularity)
    \hline\rule{0pt}{12pt}
    \multirow{4}{*}{\rotatebox[origin=c]{360}{\footnotesize $s_p$}} &
    \multirow{2}{*}{\vtop{\hbox{\strut Ext.-}\hbox{\strut Baseline}}}
        & $|c|$      &0.736$\pm$0.035     &0.061$\pm$0.028     \\[2pt]
        & & Age &0.756$\pm$0.023     &0.008$\pm$0.026     \\[2pt]

    \cdashline{2-5} \rule{0pt}{12pt}
    %\cline{2-5}\cline{2-5}\rule{0pt}{12pt}
    & \multirow{1}{*}{$BP$}
        & All &   0.628$\pm$0.038     &0.32$\pm$0.046     \\[2pt]
    % end of $BP$ models - moving to $DP$
    %\cdashline{2-5}
    & \multirow{1}{*}{$DP$}
        & All &0.593$\pm$0.057     &0.421$\pm$0.038     \\[2pt]
    \hline\rule{0pt}{12pt}
    
    % next gold label (diameter)
    \multirow{4}{*}{\rotatebox[origin=c]{360}{$s_d$}} &
    \multirow{2}{*}{\vtop{\hbox{\strut Ext.-}\hbox{\strut Baseline}}}
        & $|c|$ &0.884$\pm$0.031     &-0.002$\pm$0.021     \\[2pt]
        & & Age &0.869$\pm$0.047     &0.032$\pm$0.068     \\[2pt]
    \cdashline{2-5} \rule{0pt}{12pt}
    %\cline{2-5}\cline{2-5}\rule{0pt}{12pt}
    & \multirow{1}{*}{$BP$}
        & BOW &0.744$\pm$0.047     &0.294$\pm$0.071     \\[2pt]
    % end of $BP$ models - moving to $DP$
    %\cdashline{2-5}
    & \multirow{1}{*}{$DP$}
        & All &0.727$\pm$0.066     &0.348$\pm$0.039     \\[2pt]
    \hline\rule{0pt}{12pt}

    % last gold label (pixels + complexity)
        \multirow{4}{*}{\rotatebox[origin=c]{360}{\footnotesize $s_H$}} &
    \multirow{2}{*}{\vtop{\hbox{\strut Ext.-}\hbox{\strut Baseline}}}
        & $|c|$          &0.732$\pm$0.037     &0.044$\pm$0.036     \\[2pt]
        & & Age   &0.75$\pm$0.028     &-0.003$\pm$0.016     \\[2pt]
    \cdashline{2-5} \rule{0pt}{12pt}
    %\cdashline{2-5} \rule{0pt}{12pt}
    & \multirow{1}{*}{$BP$}
        & All &0.646$\pm$0.037     &0.255$\pm$0.037     \\[2pt]
    % end of $BP$ models - moving to $DP$
    %\cdashline{2-5}
    & \multirow{1}{*}{$DP$}
        & All &0.634$\pm$0.051     &0.327$\pm$0.046    \\[2pt] 
        
    %\hline\rule{0pt}{12pt}
    \end{tabular}
}
\vspace{8pt}
\caption{\footnotesize Regression results. \emph{Succ.}: the success-definition to be predicted. \emph{External (Ext.) Baseline}: univariate linear model. $\downarrow$ ($\uparrow$) indicates that a lower (higher) value is better. \emph{Adj.}: Adjusted.}
\label{table:reg_res}
\end{table}

\subsection{Classification Results}
\label{subsec:classification_results}
The best classification results were obtained by the GBT classifier using all types of features (textual, meta, and network) derived from the DP dataset: an average F1-score of 0.695 and AUC of 0.694, over 5-folds. These results significantly outperform the best baseline (F1-score of 0.53 and AUC of 0.55). 

Using the $BP$ data the top-performing model achieves an F1 score of 0.647 and an AUC of 0.645. While these results are inferior to the results obtained based on the DP data, they are still significantly better than the baseline results.

\subsection{Regression Results}
\label{subsec:regression_results}
Table \ref{table:reg_res} presents regression results in a 5-fold cross-validation setting. Due to space constraints, we report results only for the univariate baseline models and for the best-performing feature set for each label type and dataset. 

Using either $BP$ or $DP$ dataset significantly outperforms both baselines in all four label types. %\oren{but what model? GBT?} 
The results obtained over $DP$ consistently outperform those obtained over $BP$. We further discuss this trend in Section \ref{sec:discussion}. 

\subsection{Analysis and Social Interpretation}
\label{subsec:social_insights}
\begin{figure*}[t]
  \centering
    \begin{tabular}{c@{}}
        \includegraphics[scale=0.1]{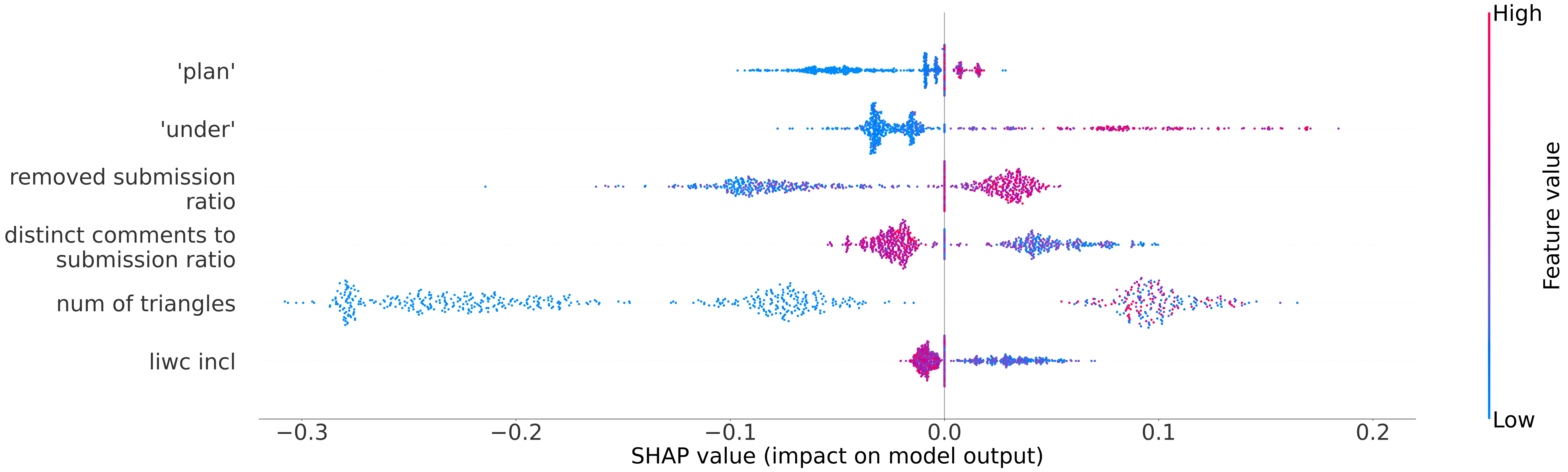}
    \end{tabular}
  \caption{SHAP values for the \textcolor{black}{six} prominent features in modeling the $s_{p}$ label while using the $DP$ dataset.}
  \label{fig:shap_values}
\end{figure*}

The SHAP explanatory toolkit \cite{lundberg2017unified} allows quantifying the impact of specific features on the prediction. A high (low) SHAP value indicates the feature's positive (negative) impact on the prediction for a specific instance (community). We use the SHAP aggregate values to derive social insights.
SHAP values of \textcolor{black}{six prominent features (two word-tokens, two community meta features, one network feature, and one LIWC feature)} are presented in Figure \ref{fig:shap_values}. Higher X-axis values indicate a positive contribution to the model's prediction. The color corresponds to the actual value of the feature. \textcolor{black}{The analysis we provide was done on the DP setting.}

\paragraph{Planning, alerting, engaging} We observe that high values of the word-feature \emph{plan} is correlated with a positive SHAP value. 
We manually verified that the frequent use of the word is often used in the context of strategic planning of the community's action. Similarly, the word-feature \emph{under} is used to alert the community members as in ``we are [our logo is] under attack''. \textcolor{black}{On the other hand, we observe a \emph{negative} correlation between the LIWC category `incl' (Inclusive\footnote{\textcolor{black}{Examples of words in this category:  \emph{with}, \emph{together} and \emph{plus}.}}) and success in the \rp~experiment.}

Interestingly, we observe that a low \emph{distinct comments to submission ratio} (the number of \emph{users} responding to a submission, not to confuse with \emph{comments to submission ratio}) has a positive effect on the prediction score. Our interpretation suggests that successful campaigns allow focused discussions between community members: comments and discussions are encouraged as long as a discussion thread does not lose focus. Too many \emph{members} commenting on a submission (high feature values) may result in stagnation that harms coordination. \textcolor{black}{Furthermore, high values of the \emph{removed submission ratio} is positively associated with success in the game. Removal of submission in Reddit is allowed by the author himself or by the moderators of the community (i.e., a small set of users that manage the community). 
These features suggest that (self) moderation is important in large-scale distributed campaigns.}

The SHAP values for \emph{num of triangles} provide a complementary perspective: a denser network is related to better performance -- \textcolor{black}{reinforcing the sociology scholarship asserting that high clustering facilitates trust and high social capital \cite{coleman1988social}}. Combined with our interpretation of the SHAP analysis of the \emph{distinct comments to submission ratio} it suggests that successful communities have their members engaged efficiently in focused discussions, rather than being verbose, creating distractions. This interpretation provides additional nuance to well-established theory, e.g., \citet{backstrom2006group, cunha2019all}.

%Figure \ref{fig:shap_values} highlights that successful communities tend to have \emph{less} responders per submission. %The top 10\% most successful communities had, on average, only ?? responders per submission, while the top 10\% most unsuccessful had ?? on average.
%We hypothesize that this indicates that communities that operate through explicit guidelines/commands during the \rp~experiment are more likely to succeed. Figure \ref{fig:shap_values} demonstrates that communities with a \emph{high (low)} number of triangles are more likely to succeed. 

% The SHAP figure

% a figure showing LOL artwork and its SHAP values
\ignore{
    \begin{figure}
    \normalsize
        \centering
        \subfloat[\label{subfig:lol_artwork}]{{\includegraphics[scale=0.45]{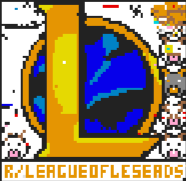}}}
        \quad
        \subfloat[ 
        \label{subfig:shap_feature_importance}]{{\includegraphics[scale=0.12]{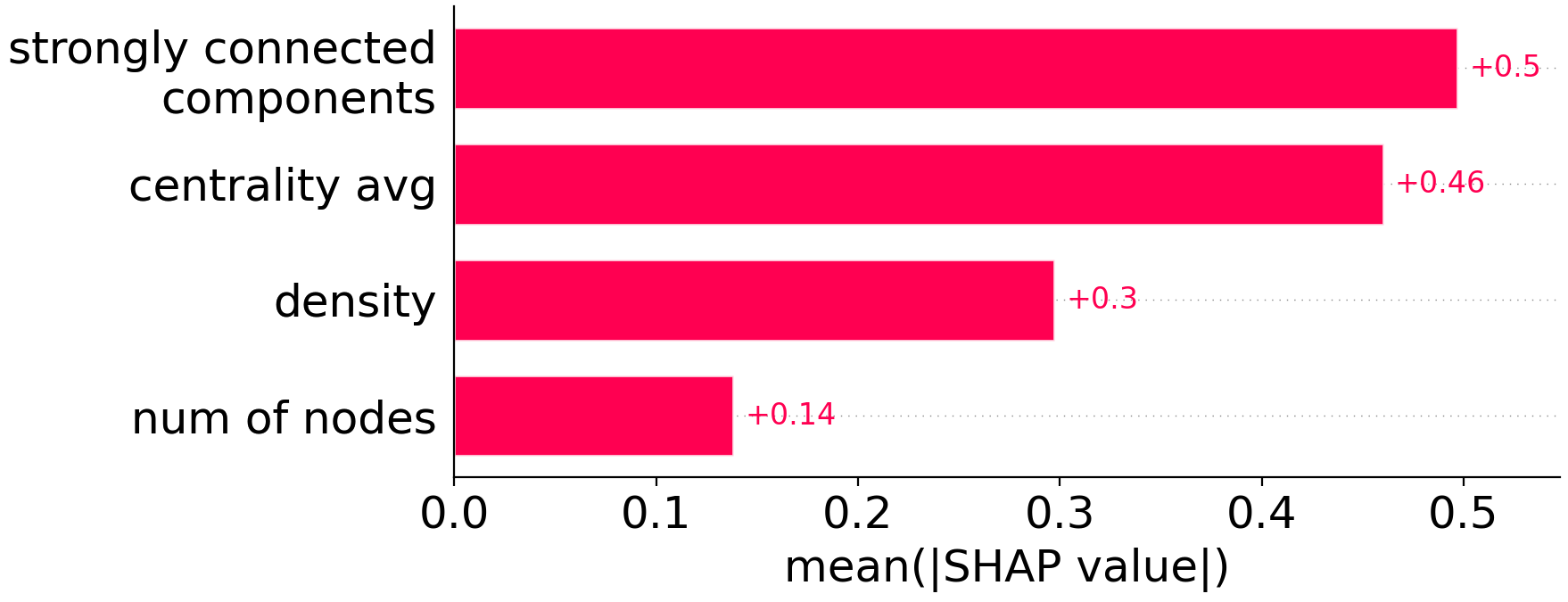}}}
    \caption{(a) LeaguOfLegends (LOL) artwork. The artwork consists of 7114 pixels. Its' $s_{|c|}$ value is relatively low due to the extremely high number of members (1.96M) in the community. On the other hand, its' $s_{H}$ value is relatively high due to the high entropy of the artwork.
    (b) SHAP feature importance. The four most dominant network features to predict the LeaguOfLegends community success level. The $s_{|c|}$ label is used in this analysis.}
    \end{figure}
}

% the LOL artwork
\begin{figure}[t]
    \centering
    \begin{tabular}{c@{}}
        \includegraphics[scale=0.98]{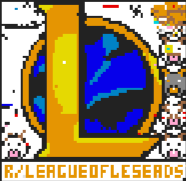}
    \end{tabular}
    \caption{\small LeaguOfLegends artwork. The artwork is located at the center of the canvas, left of the France flag.}
    \label{subfig:lol_artwork}
\end{figure}

\paragraph{Differences between definitions of success} Previous work by \citet{cunha2019all} identified four different measures of long-term success. In Section \ref{sec:definitions_of_success} we presented and motivated five measures of success in concrete campaigns. In this part, we explore the differences  between success measures and the correlated factors by focusing on communities for which performance differs radically across success measures\footnote{Due to space constraints our we provide analysis only for the $s_d$ (logo diameter) and $s_H$ (complexity) success measures.}. 
For example, the logo produced by the \emph{League of Legends} (LoL) gaming community consists of 7114 pixels (see Figure \ref{subfig:lol_artwork}). Given that it is one of the largest communities in the corpus ($\sim$ two million members), $s_{|c|}(LoL)$ success level is ranked 644 out of the 666 surviving communities (i.e., that left a recognizable mark on the canvas). However, considering the complexity of the logo, $s_H(LoL)$ ranks the community at 14/666.

In total, we identified 134 communities in which their success rank is radically different according to different success measures (top quartile in one measure and in the bottom quartile in another). We denote this set of communities $c^\Delta$.

Table \ref{table:success_measures_analysis} presents the average ablation RMSE for $s_{|c|}$ and $s_H$ on all communities in $C^\Delta$. 
It is evident that different feature types play a more/less significant role, depending on the definition of success. For example, the Meta and Graph2Vec features better predict success when accounting for community size. In contrast, BOW and Com2Vec features are helpful where the focus is on the complexity of the goal. 

Using SHAP analysis to recover the role of specific features in predicting $s_{|c|}$, we find that the number of community members, the number of comments posted, and the average upvote score are the most important meta features; the average centrality, the density of the network, and the number of triangles are found to be the best predictors among the network features. \textcolor{black}{%This result reflects two core concepts suggested by \citet{mcmillan1986sense} -- \emph{membership} and \emph{influence}. For example, a culture of positive feedback, evidenced by high upvote scores, is correlated with community success. 
This is inline with the findings of \citet{cunha2016effect} and \citet{cheng2014community} regarding the importance of engagement and positive feedback.}

\begin{table}
\centering
{
    \begin{tabular}{c@{\quad}|c@{\quad}c@{\quad}}
      Feature Type & $s_{|c|}$ & $s_H$ \\[2pt]\hline
      Meta & {\bf 0.438} & 0.51 \\[2pt]
      Network &  0.444 & 0.525\\[2pt]
      LIWC & 0.532 & 0.549\\[2pt]
      BOW &  0.529 & {\bf 0.489}\\[2pt]
      Com2Vec &  0.478 & {\bf 0.5}\\[2pt]
      SNAP &  0.501 & 0.55\\[2pt]
      Graph2Vec &  {\bf 0.427} & 0.521\\[2pt]
      \cdashline{1-3} \rule{0pt}{12pt}
      All &  0.436 & 0.504\\[2pt]
    \end{tabular}
\caption{\small Average RMSE values of $s_{|c|}$ and $s_H$ success prediction for communities in $C^\Delta$. Best performing feature types are marked in boldface.}
\label{table:success_measures_analysis}
}
\end{table}

% the (nice) table was removed due to lack of space :(
\ignore{
    \begin{figure*}
        \centering
        \begin{tabular}{c@{}}
            \includegraphics[scale=0.15]{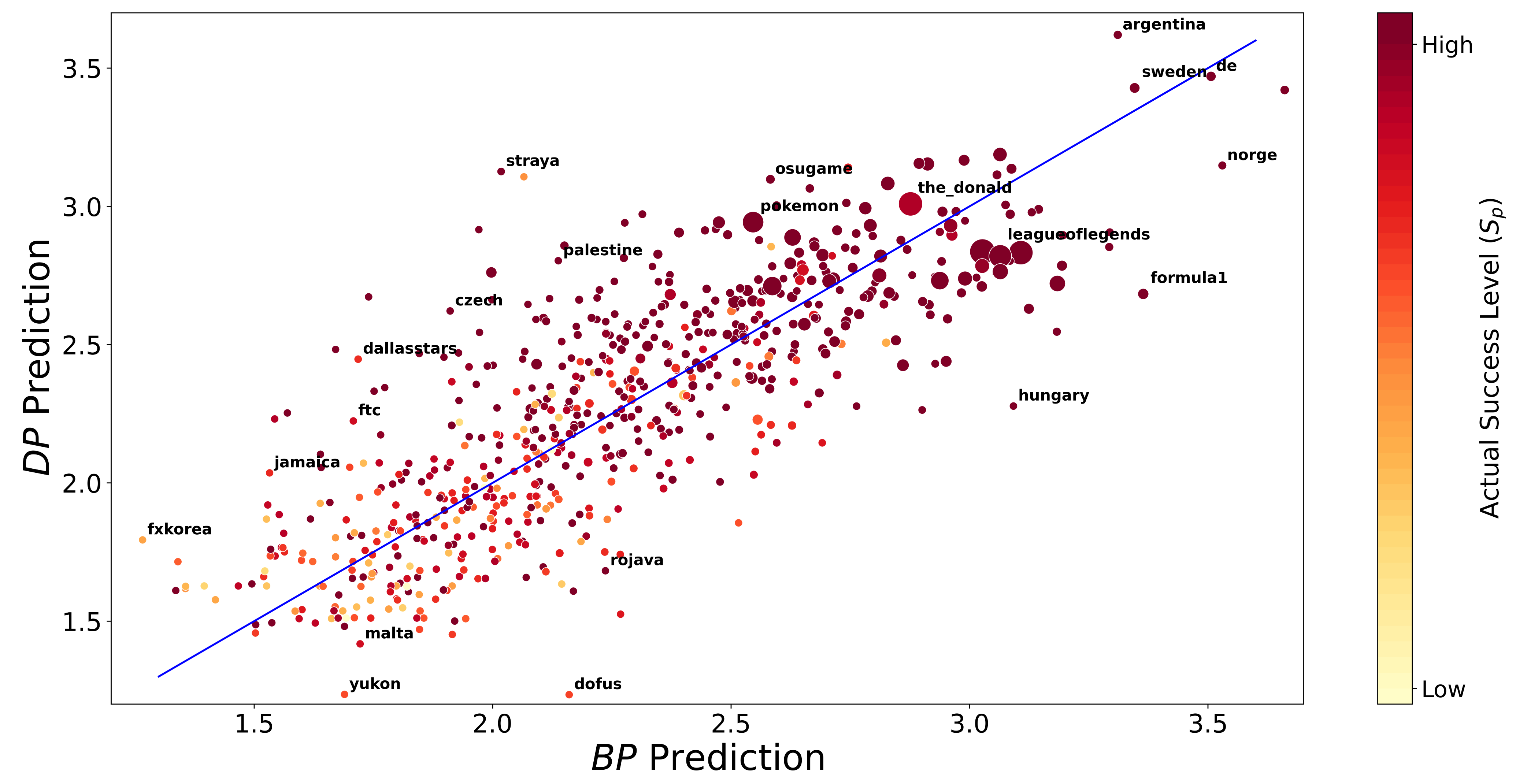}
        \end{tabular}
        \caption{\small \textcolor{black}{$BP$ VS. $DP$ predictions. Each point represents a community in \rp~. The size of each point represents the community size. Colors emphasize the success level of the community, using the $S_p$ measure. The 45$^{\circ}$ blue line is drawn for better interpretation of the figure.}}
        \label{fig:bp_vs_dp}
    \end{figure*}
}

\paragraph{\textcolor{black}{Differences between $BP$ and $DP$}} \textcolor{black}{We further analyze the differences between the community signal obtained prior to the experiment and the signal obtained during the experiment. 
%We compare them on different levels. 
Due space constraints, we present the analysis only over the $S_p$ success label .\footnote{\textcolor{black}{The $S_p$ achieves the best $R^2$ over all the labels, see Table \ref{table:reg_res}.}} A comparison between $BP$ and $DP$ is provided in Table \ref{table:bp_vs_dp}. \footnote{\textcolor{black}{Note that the SNAP and Com2Vec features types are not included since they are obtained from  external sources that do not provide different representations for $BP$ and $DP$.}} 
We observe that the $DP$ model better predicts success over all types of features. In both cases, the model that combines all types of features performs best.}

\textcolor{black}{To find further differences between $BP$ and $DP$, we analyze the feature importance distribution (by SHAP values) of the models that combine \emph{all} feature types. We find that in both models the `BOW' features contribute the most (39.8\% and 60.5\% in the $DP$ and $BP$ models respectively). However, the `Network' features contribute 30.6\% of the features' importance in the $DP$ model compared to 14.7\% in the $BP$ model. We also observe a stronger contribution of the LIWC features in the $DP$ model (16.2\% VS. 12.8\%).}

\textcolor{black}{We further contrasted the predictions of the models on the same community, finding the cases in which the models' predictions differ significantly. Such cases highlight \emph{outlier communities} that over (or under) perform during the \rp~experiment. Two such outlier communities are r/osugame (see Figure \ref{fig:canvas_evolution_conflict_regions}, last row) and r/straya (Australia). In both communities, the `Network' features were significantly more dominant in the $DP$ model compared to the importance of the `Network' features in the $BP$ model.}

\textcolor{black}{Interestingly, national/geographical communities (e.g., r/straya) are over-represented in the set of outlier communities. Out of the top 100 outlier communities, 23\% are national/geographical communities, while  only 10.6\% (106 out of 997) of the participating communities  are national/geographical communities. We hypothesize that such an over-representation is due to the nature of the \rp~settings. Flags (and other national symbols) are recognizable by all community users, uncontroversial among most members, and relatively easy to draw. Moreover, it is well established that national identity is one of the stronger totems of personal identity, having individuals unite, fight and protect national symbols \cite{mudde2007populist,reicher2000self,smith2013nationalism,jaskulowski2016magic}.}

\begin{table}
\centering
\small
{
    \begin{tabular}{c@{\quad}|c@{\quad}c@{\quad}|c@{\quad}}
    Feature Type & \boldmath{$BP$} & \boldmath{$DP$} & {Cor.}\\[2pt]\hline
    Meta & 0.26 & 0.33 & 0.81\\[2pt]
    Network & 0.27 & 0.328 & 0.83\\[2pt]
    LIWC & 0.172 & 0.269 & 0.59\\[2pt]
    BOW & 0.301 & 0.364 & 0.72\\[2pt]
    Graph2Vec & 0.25 & 0.29 & 0.78\\[2pt]
    \cdashline{1-4} \rule{0pt}{12pt}
    All & 0.322 & 0.421 & 0.8\\[2pt]
    \end{tabular}
\caption{\small $R^2$ results (higher is better) for the location popularity success measure ($S_p$) using $BP$ and $DP$ datasets. Cor. represent the Pearson correlation between the two predictions.}
\label{table:bp_vs_dp}
}
\end{table}

\section{Discussion}
\label{sec:discussion}
\paragraph{The limited performance of LLMs} State-of-the-Art results in many prediction tasks are often achieved by fine-tuning LLMs. We have experimented with a number of LLMs, including distillBERT \cite{sanh2019distilbert} and the Longformer \cite{beltagy2020longformer} which is more adequate to handle longer sequences of texts. The performance of the LLMs were disappointing\footnote{For example, in the binary classification task (``leaving a mark'') the BERT model achieved an average F1-score of 0.627 using the $BP$ dataset, compared to 0.647 by the GBT.}.
We attribute the modest performance of LLMs on the task and data at hand to two factors: First, the number of instances (communities) is relatively small, which may not be enough for training large models. Second, LLMs capture the topic and semantics of the texts, but these signals are not as important as the community structure and the community dynamics. 

\paragraph{Success measures and the community objective} We observe a significant difference in the success (gold labels and predicted labels) with respect to the different measures of success.  %This differences are not necessarily reflected in terms of the label values.
It is interesting to note that in practice, the $s_{d}(i)$ and $s_{H}(i)$ values depend on the explicit objective the community members aim to achieve as they decide on the artwork's shape and complexity. On the other hand, the $s_{|c|}(i)$ is not directly controlled by the $c^i$ since the size of each community is mostly fixed prior to \rp~(new users cannot join, although registered users can migrate between communities). Finally, $s_{p}(i)$ is controlled to some degree by $c^i$: some communities deliberately choose to operate in regions of high demand (e.g., the U.S. flag in the center of the canvas), while other communities (e.g., r/osuGame) operated at the periphery of the canvas just to be repeatedly attacked by TBV. These clashes made the OSU location the most popular area in term of pixel changes.

\paragraph{\textcolor{black}{Generalizability}} \textcolor{black}{We use the \rp~experiment to model success levels of online communities. The unique setting allows us to consider multiple ways to quantify success. On the one hand, the set of explanatory features and success measures are specific to \rp, while on the other hand, we use general \emph{concepts} that can be used in other experimental settings. For example, the entropy of the artwork and the location on the canvas are \rp-specific, but the complexity of the group's objective and the opposition it faces are general. Similarly, a specific word-token can have a high SHAP value in this context, but using word tokens, community structure, and other features are general enough and could be used for many modeling tasks.} %This is weakly relevant for a moderator that estimates the success of communities in an online political campaign. However, taking into account the complexity of the ideas disseminated by the community is highly important and relevant.}

\paragraph{\textcolor{black}{Limitations}} \textcolor{black}{The nature of the Reddit platform and the \rp~experiment attracted specific communities and demographics -- a few million users organized in about a thousands of communities make only a small part of the users and the communities\footnote{There are $\sim 1.7B$ registered users and $\sim 1.2M$ communities, though many are inactive.} on Reddit. Our modeling and analysis only include those that participated.}

\textcolor{black}{In this work, we propose different success measures that rely on previous studies as well as the nature of the \rp~experiment. However, quantifying success according to the \emph{objective measures} internally defined per community would be a more suitable choice (e.g., blocking another competing community). Recovering goals this specific is extremely challenging and may not even have any clear indication in the data.}

\section{Related Work}
\label{sec:related_work}
\paragraph{Community dynamics and collaborative action}The behaviors, norms, and dynamics of human communities are at the core of the social science research \cite{lewin1947frontiers,lewin1948resolving,lewin1947group}. Naturally, in the last decade, much of the research has been geared towards online communities over social platforms, e.g., \cite{lazer2009life,zhang2017community,mensah2020characterizing}.

Reddit data have been used extensively to study various aspects of the organization, development, evolution, and behavior of online communities. A general overview of the study of Reddit communities is provided by \citet{medvedev2017anatomy}.  
While in our work, we study hundreds of communities, other works focus on a \emph{single} community, presenting its uniqueness and norms \cite{jones2019r, august2020explain, britt2021oral}.

Evolving community behaviors, the effect of moderation on Reddit communities, and the different factors that cause a community to evolve are studied in a battery of studies \cite{weninger2013exploration,choi2015characterizing,stoddard2015popularity,cunha2016effect,panek2018effects,fiesler2018reddit,rappaz2018latent,mensah2020characterizing} to mention just a few. %removed to due lack of space: de2014mental, newell2016user
These works address various aspects of community organization as an interest group, the dealings with topics of interest, and the inherent tension between anonymity and identity. Recent works study the structure and other characteristics (e.g., loyalty) of Reddit communities \cite{zhang2017community, hamilton2017loyalty,kumar2018community, zhou2020condolences, massachs2020roots}. \citet{zhang2017community} suggests a new representation of communities through 'distinctiveness' and 'dynamicity' dimensions. The authors emphasize that these representations reflect different user engagement measures (e.g., retention rate). \citet{kumar2018community} suggests a novel way to model conflicts between online communities. Their approach integrates textual data with communities' meta-features. This methodology is similar in a way to the methodology we use to combine different representations of communities. \citet{datta2019extracting} expand this work and study the landscape of conflicts among communities on Reddit.

\paragraph{Success of communities} In this work, we model community \emph{success}. A series of studies tackle this topic using multiple definitions of success \cite{kairam2012life,tan2018tracing,cunha2019all}. These definitions rely on measures that are associated with the \emph{activity} of a community. E.g., the number of posts generated and growth rate, and the members' retention.

Our \emph{success} definition is purely different from these works. We define and model success quantitatively, based on the performance of each community in a naturally occurring-large scale experiment.

\citet{cunha2019all} identify four success measures associated with communities and analyze their relationship. %\oren{need to list (or maybe list and discuss in the discussion?)}
They conclude that success is multi-faceted and can hardly be measured nor predicted by a single measurement. Their work and approach inspire our work. We hypothesize that success in the \rp~experiment has to be measured using multiple measurements, each capturing a different facet of success. We also assume that a unique predictive model should be fitted per success measure.

\paragraph{The r/place experiment} Studies that utilize the \rp~experiment data are still scarce. \citet{muller2018compression} study the experiment from a perspective of how artworks evolve over time and their correlation with the canvas density. \citet{rappaz2018latent} and \citet{armstrong2018coordination} introduce an analysis of the latent patterns of collaboration between individuals. Conflicts between communities during the \rp~experiment are studied by \cite{vachher2020understanding}, which also releases a dataset of conflict regions and the communities involved in each. None of these works tackle the prediction tasks we propose. In addition, they fundamentally differ from our work since they focus on modeling \emph{individual redditors} while we focus on the \emph{community} level, and they only use the \rp~pixel allocations data while we combine \emph{multiple types of features} (language, community structure, user dynamics, etc.).

\citet{litherland2021instruction} introduced a framework to analyze specific communities that draw artwork in \rp. They study the evolution of visual artifacts and social artifacts during the experiment. However, they use limited data for the analysis (only structural) and focus on a single community (the Mona Lisa painting).

Community engagement in large-scale distributed campaigns was recently studied by \citet{israeli2022must}. This work is closely related to our work. Both works focus on communities rather than individual users and use a similar computational approach involving the integration of multiple feature types into the prediction model. However, our work differs from \shortcite{israeli2022must} in three key aspects: (i) We define a different research question -- rather than focusing on the (non)participation of a community in \rp, we focus on the \emph{success level} of participating communities, (ii) While \shortcite{israeli2022must} use only data that was generated before \rp, we also examine the signal produced \emph{while} the experiment. We compared the performance of the models using data generated before and during the experiment, and (iii) We follow  \citet{kumar2019predicting} and use community embeddings, together with `naive' representations based on a predefined set of features (e.g., centrality).%(e.g., modularity, centrality, and degree distribution).

\section{Conclusions and Future Work}
\label{sec:closure}
We study how community structure, language, internal norms, and other characteristics can be used to predict the success level of a community in large-scale distributed campaigns. Specifically, we predict how well a Reddit community performs in the \rp~experiment.
We argue that success can be defined in a number of ways that are not well correlated. Defining a number of success measures we experimented with a high number of representations per community (e.g., language and network) -- calculated before and during the \rp~experiment.

We found that the data collected \emph{during} the experiment are more effective than the data collected \emph{before} the experiment, \textcolor{black}{overall and for each feature type separately}. We find that certain words, such as \emph{plan} and \emph{under} are highly correlated with the success level in \rp. We also find structural characteristics such as the number of triangles and the network density that are positively correlated with success in the game. \textcolor{black}{We find that communities conducting more focused discussions have a better success rate. Finally, relying on a novel comparison between two of our models, we find that the success level of national/geographical communities (e.g., r/straya) is not well predicted by the models, suggesting that these communities have a unique behavior while engaging in the \rp~experiment.}

Future work takes \textcolor{black}{two} trajectories: (i) Model the 2022 \rp~experiment\footnote{A new version of the \rp~experiment that attracted 16M redditors but had a very different setting than the 2017 experiment.} and (ii) Model the behavior and collaboration between \emph{users} in the \rp~experiment.%\textcolor{black}{and (iii) Model the behavior and success of national/geographical communities in the \rp~experiment.}

\paragraph{Broader perspective: ethics} One primary objective of this work is to better understand the factors that contribute to a successful undertaking of a campaign by a community. The insights derived from the predictive models and from the analysis of the results can help communities to coordinate in order to promote a cause or mitigate adversarial campaigns. On the other hand, these insights could also be utilized in designing more efficient adversarial campaigns such as election interference, increasing polarization, or promoting fake news or anti-scientific sentiment. 

\bibliography{icwsm_2023_bib_file}
\end{document}